\documentclass{ieeeaccess}

\usepackage{cite}
\usepackage{amsmath,amssymb,amsfonts}
\usepackage{algorithmic}
\usepackage{graphicx}
\usepackage{textcomp}

\usepackage{subcaption}
\usepackage{xcolor, colortbl}
\usepackage{supertabular,booktabs,calc}
\usepackage{multirow}
\usepackage{caption}

\usepackage{placeins}
\usepackage{lscape}
\usepackage{array}

\def\BibTeX{{\rm B\kern-.05em{\sc i\kern-.025em b}\kern-.08em
    T\kern-.1667em\lower.7ex\hbox{E}\kern-.125emX}}

\newcolumntype{P}[1]{>{\centering\arraybackslash}m{#1 -1.8\tabcolsep}}

\newenvironment{IEEEbiographynophotoAlt}[1]{%
\normalfont\biographyfont\interlinepenalty500%
\vskip 4\baselineskip plus 1fil minus 0\baselineskip%
\parskip=0pt\par%
\noindent{\biographyheadfont\uppercase{#1}\ }}{\relax\par\normalfont}

\let\keptmaketitle\maketitle
\usepackage{hyperref}
\newcommand{\orcid}[1]{\href{https://orcid.org/#1}{\includegraphics[scale=0.7]{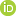}}}

\usepackage{soul}
\soulregister\cite7
\soulregister\ref7
\soulregister\pageref7

\let\maketitle\keptmaketitle

\begin{document}
\history{Date of publication xxxx 00, 0000, date of current version xxxx 00, 0000.}
\doi{10.1109/ACCESS.2017.DOI}

\title{A Taxonomy of Network Threats and the Effect of Current Datasets on Intrusion Detection Systems}
\author{
\uppercase{Hanan~Hindy}\orcid{0000-0002-5195-8193}\authorrefmark{1}, \IEEEmembership{Member, IEEE},
\uppercase{David~Brosset\orcid{0000-0002-9677-1445}\authorrefmark{2}, 
Ethan~Bayne\orcid{0000-0003-1853-2921}\authorrefmark{1}, 
Amar~Seeam\orcid{0000-0002-8393-3214}\authorrefmark{3}, \IEEEmembership{Member, IEEE}, 
Christos~Tachtatzis\orcid{0000-0001-9150-6805}\authorrefmark{4}, \IEEEmembership{Senior Member, IEEE}, 
Robert~Atkinson\orcid{0000-0002-6206-2229}\authorrefmark{4}, \IEEEmembership{Senior Member, IEEE} and Xavier~Bellekens}\orcid{0000-0003-1849-5788}\authorrefmark{1,4},
\IEEEmembership{Member, IEEE}.}
\address[1]{Division of Cyber Security, Abertay University, Dundee, Scotland, UK}
\address[2]{Naval Academy Research Institute, France}
\address[3]{Department of Computer Science, Middlesex University, Mauritius}
\address[4]{EEE Department, University of Strathclyde, Glasgow, Scotland, UK}

\markboth
{H. Hindy \headeretal: A Taxonomy of Network Threats and the Effect of Current Datasets on Intrusion Detection Systems}
{H. Hindy \headeretal: A Taxonomy of Network Threats and the Effect of Current Datasets on Intrusion Detection Systems}

\corresp{Corresponding author: Hanan Hindy (e-mail: hananhindy@ieee.org).}

\begin{abstract}
As the world moves towards being increasingly dependent on computers and automation, building secure applications, systems and networks are some of the main challenges faced in the current decade. 
The number of threats that individuals and businesses face is rising exponentially due to the increasing complexity of networks and services of modern networks. 
To alleviate the impact of these threats, researchers have proposed numerous solutions for anomaly detection; however, current tools often fail to adapt to ever-changing architectures, associated threats and zero-day attacks.
This manuscript aims to pinpoint research gaps and shortcomings of current datasets, their impact on building Network Intrusion Detection Systems~(NIDS) and the growing number of sophisticated threats. 
To this end, this manuscript provides researchers with two key pieces of information; a survey of prominent datasets, analyzing their use and impact on the development of the past decade's Intrusion Detection Systems~(IDS) and a taxonomy of network threats and associated tools to carry out these attacks.  
The manuscript highlights that current IDS research covers only 33.3\% of our threat taxonomy. Current datasets demonstrate a clear lack of real-network threats, attack representation and include a large number of deprecated threats, which together limit the detection accuracy of current machine learning IDS approaches. The unique combination of the taxonomy and the analysis of the datasets provided in this manuscript aims to improve the creation of datasets and the collection of real-world data. As a result, this will improve the efficiency of the next generation IDS and reflect network threats more accurately within new datasets.
\end{abstract}


\titlepgskip=-15pt

\maketitle
\section{Introduction}
\PARstart{T}{he}  world is becoming more dependent on connected devices, actuators and sensors, regulating the lives of millions of people. Furthermore, sensor data is expected to increase by around 13\%, reaching 35\% of overall data communication in 2020, reaching a peak of 50 billion connected devices and an increased monthly Internet traffic volume reaching 30~GB on average per capita compared to around 10~GB in 2016~\cite{RefWorks:doc:5ae8785ae4b0c16216f56d10}.
While each device in an Internet of Things~(IoT) system exchanges data, associated services often provide interfaces to interact with the collected data, often increasing the attack surface. Therefore, it is crucial to build robust tools to defend networks against security threats in modern IoT networks. Current detection tools are often based on outdated datasets that do not reflect the reality of recent/modern network attacks, rendering Intrusion Detection Systems (IDS) ineffective against new threats and zero-days. 
To the best knowledge of the authors, there are currently no manuscripts that analyze the shortcomings of available networking datasets, nor provide a taxonomy of the current network threats and the associated tools used to carry out these attacks.

The contributions of this research are threefold:
\begin{itemize}
\item An evaluation of the limitations of the available network-based datasets and their impact on the development of IDSs
\item A review of the last decade's research on NIDS
\item A Threat taxonomy is presented, and categorized by:
    \begin{itemize}
      \item The Threat Sources
      \item The Open Systems Interconnection (OSI) Layer
      \item Active or Passive modes
    \end{itemize}
\end{itemize}

The evaluation of current network-based datasets provides researchers with an insight of the shortcomings of the datasets presented when used for training against real-world threats.  A threat taxonomy is derived from the datasets and current real-world networking threats. This taxonomy serves two purposes--- firstly, it strengthens our argument on the shortcomings of currently available datasets, but most importantly, it provides researchers with the ability to  identify threats and tools underrepresented in currently available datasets. To facilitate this endeavor, we further map the current threats with their associated tools, which in turn can be used by research to create new datasets. 

The rest of the paper is organized as follows; 
Section~\ref{sec:IDS} depicts the main differences between IDSs, the metrics to consider for their evaluation and the role of feature selection in building IDSs. Section~\ref{sec:recent-ids} reviews IDSs of the past decade and their individual contributions are assessed. This section also evaluates the drawbacks and limitations of the available datasets. Section~\ref{sec:Threats} provides the threat taxonomy. Section~\ref{sec:challenges} summarizes the challenges presented in this work and provides recommendations. Finally, the paper is concluded in Section~\ref{sec:Conclusion}.
\section{Background}
\label{sec:IDS}

\subsection{Intrusion Detection Systems}
IDSs are defined as systems built to monitor and analyze network traffic and/or systems to detect anomalies, intrusions or privacy violations.
When an intrusion is detected, an IDS is expected to (a)~log the information related to the intrusion, (b)~trigger alerts and (c)~take mitigation and corrective actions~\cite{RefWorks:doc:5ae87184e4b0a553e076cb0a}.

IDS can either be Host Intrusion Detection System~(HIDS) or Network Intrusion Detection System~(NIDS). HIDS is responsible for monitoring a system internally, having access to log files, users' activities, etc. While NIDS analyses incoming and outgoing communication between network nodes.

IDSs differ based on their detection method. Signature-based IDSs were the first to be developed. Accurate signatures are built from prior detected attacks. The main advantage of this method is the high accuracy of detecting known attacks. Signature-based IDS is, however, unable to detect zero-days, metamorphic and polymorphic threats~\cite{bellekens2014glop}. 
The second method, Anomaly-based detection, depends on identifying patterns and comparing them to normal traffic patterns. This method requires the system to be trained prior to deployment. The accuracy of anomaly-based systems against zero-days, metamorphic and polymorphic threats is better when compared to signature-based IDS. However, the false positive rate of anomaly-based detection is often higher. It is important to mention that benign/normal traffic patterns alone are not sufficient to detect attacks. For this reason, the features used to represent network traffic play an essential role in traffic representation.

Intrusion detection (both signature-based and anomaly-based) can be done on a stateless (per packet) or stateful (per flow) basis. Most recent IDSs are stateful, as the flow provides ``context'', while packet analysis  (stateless) does not provide this context. It is the responsibility of the researcher to decide which method is best suited for their application.

Anomaly-based IDS can be classified into subcategories based on the training method used. These categories are statistical, knowledge-based and Machine Learning~(ML) based. Statistical includes univariate, multivariate and time series. Knowledge-based uses finite state machines and rules like case-based, N-based, expert systems and descriptor languages. Buczak and Guven~\cite{7307098} provide recommendations on choosing the ML/Deep Learning~(DL) algorithms based on the problem intended to be solved. Algorithms include Artificial Neural Networks~(ANN), clustering, Genetic Algorithms~(GA), etc. 
Specification-based combines the strength of both signature and anomaly based to form a hybrid model.

Owezarski~\textit{et al.}~\cite{owezarski2008laasnetexp} summarize the approaches to validate networking models, which applies to IDS, into four categories; mathematical models, simulation, emulation and real experiments. Each of these approaches has their own pros and cons as discussed by Behal and Kumar~\cite{BEHAL20167}.

\subsubsection{Metrics for IDS Evaluation}
In order for an IDS to be considered effective, high detection rate and low false positive rate are key aspects to consider. Multiple metrics could be used for an IDS evaluation. These metrics are discussed subsequently showing the significance and purpose of each. It is important to mention that depending only on detection rate as the only evaluation metric doesn't reflect an IDS performance.

Other important evaluation factors including the transparency and safety of the overall system, memory requirements, power consumption and throughput should  be considered.
Moreover,~\cite{Axelsson:2000:BFD:357830.357849} adds to the aforementioned requirements, ease of use, interoperability, transparency and collaboration.

\noindent
IDS accuracy can be defined in terms of:
\begin{itemize}
\item True Positive~(TP): Number of intrusions correctly detected
\item True Negative~(TN): Number of non-intrusions correctly detected
\item False Positive~(FP): Number of non-intrusions incorrectly detected
\item False Negative~(FN): Number of intrusions incorrectly detected
\end{itemize}

Hodo~\textit{et al.}~\cite{RefWorks:doc:5ae87684e4b0e00594a6a318}, Buse~\textit{et al.}~\cite{RefWorks:doc:5ae8697de4b066d2d904af88} and Aminanto~\textit{et al.}~\cite{RefWorks:doc:5ae87445e4b029aa3efa7c22} discuss the main metrics to consider for evaluation in their respective work. These include the overall accuracy, decision rates, precision, recall, F1 and Mcc. 

\noindent Equation~\ref{eq:OAccuracy} provides the overall accuracy. It returns the probability that an item is correctly classified by the IDS.
\begin{equation}
\label{eq:OAccuracy}
Overall Accuracy = \frac{TP + TN}{TP + TN + FP + FN }
\end{equation}

\noindent Equation~\ref{eq:detection-rate} calculates the Sensitivity, Specificity, Fallout, and Miss Rate detection rates, respectively. 
Stefan Axelsson~\cite{Axelsson:2000:BFD:357830.357849} stresses the fact that false positive rates~(false alarms) highly limit the performance of an IDS due to the "Base-rate fallacy problem".

\noindent Detection Rates:
\begin{equation} 
\label{eq:detection-rate}
\begin{split}
&Sensitivity\\&\hspace{8mm}(aka \;  Recall,\;True\;Positive\;Rate)= \frac{TP}{TP + FN} \\
&Specificity\\&(aka \; Selectivity,\;True\;Negative\;Rate) = \frac{TN}{TN + FP} \\
&Fallout\\&\hspace{19mm}(aka \; False\;Positive\;Rate) = \frac{FP}{TN + FP} \\
&Miss\;Rate\\&\hspace{18mm}(aka \; False\;Negative\;Rate) = \frac{FN}{TP + FN} 
\end{split}
\end{equation}

\noindent Equation ~\ref{eq:precision} provides the percentage of positively classified incidents that are truly positive.
\begin{equation} 
\label{eq:precision}
Precision = \frac{TP}{TP + FP}
\end{equation}
 
\noindent To visualize the performance of an IDS, i.e. the trade-off between sensitivity (true positive rate) and fallout (true negative rate), AUC~(Area Under The Curve) ROC~(Receiver Operating Characteristics) also known as Area Under the Receiver Operating Characteristics~(AUROC) curve is used~\cite{4738466, RefWorks:doc:5d3eead4e4b0f3a8f3433a51, RefWorks:doc:5d3eea07e4b0bdf80ba1e685}
\\
\noindent Equation ~\ref{eq:f1} represents the harmonic mean of precision and recall. F1 is better suited to represent the performance of an IDS, specially when dealing with imbalanced classes.
\begin{equation} 
\label{eq:f1}
F1 = \frac{2TP}{2TP + FP + FN}
\end{equation}

\noindent Equation ~\ref{eq:mcc} provides Matthews correlation coefficient. It can only be used in binary IDS in which incidents are classified as either attack or normal.
\begin{equation} 
\begin{split}
\label{eq:mcc}
Mcc &= \\ &\frac{(TP * TN)-(FP * FN)}{\sqrt{(TP+FP)(TP+FN)(TN+FP)(TN+FN)}}
\end{split}
\end{equation}

\noindent Equation~\ref{eq:geomacc} addresses the problem of calculating accuracy in imbalanced datasets. Numerous datasets have a limited number of attacks' data compared to benign traffic, hence, the geometric mean of accuracy provides a more precise metric than overall accuracy measure~\cite{kubat1997addressing}. Space-Time aware evaluation is introduced by Feargus Pendlebury~\textit{et al.}~\cite{pendlebury2018tesseract} to overcome both spatial and temporal biases. The authors introduced three constraints to be considered when splitting datasets.

\begin{flalign}
\label{eq:geomacc}
    gAcc &= \sqrt{a^{+ve}.a^{-ve}} & \nonumber \\
        &= \sqrt{Sensitivity_{(TPR)}.Specificity_{(TNR)}}
\end{flalign}

\noindent
Additionally, CPU consumption, throughput and power consumption are important metrics for evaluating IDSs. Specifically, these metrics are important for IDSs running on different hardware or with specific settings such as high-speed networks, or on hardware with limited resources.

\noindent
\subsubsection{Feature Selection and IDS}
"Feature Learning"~\cite{RefWorks:doc:5ae87445e4b029aa3efa7c22} or "Feature Engineering"~\cite{RefWorks:doc:5ae8746ae4b029aa3efa7c2b} plays an essential role in building any IDS in a way that chosen features highly affect the IDS performance. 
Features are obtained using one of three processes; construction, extraction and selection. Selection involves filter, wrapper and embedded techniques~\cite{chen2006survey}. A classification of the features used in recent datasets is provided in~\cite{7307098}.

Different features representations (i.e. abstractions)  are used to address different areas of threat detection.  Some of them could be considered na\"ive when they contain basic network information.  Others are considered rich when they represent deeper details~\cite{RefWorks:doc:5ae8746ae4b029aa3efa7c2b}. 
As highlighted by Rezaei and Liu~\cite{8713803}, there are four main categories of networking features; time series, header, payload and statistical. Unlike header and payload features, time series and statistical ones are available for both encrypted and unencrypted traffic. The authors further discuss the shortcomings of current encrypted traffic classification research. Packet-based and flow-based features have been used for intrusion detection purposes. However, with the advancement of network encryption, packet-based features are rendered impractical for complex communication networks. 

\subsection{Related Work}
In the past decade, numerous IDSs were developed and evaluated against a range of published available datasets. Diverse review and comparative studies have been published tackling the design of IDS for various applications, as well as, the machine learning techniques used to build IDS, however, the dataset challenges are not discussed.

Current Network IDS surveys often focus on a single aspect of IDS evaluation. Buczak and Guven~\cite{7307098} focus on the different ML and DL algorithms used to build IDS. They explain the different algorithms, mention their time complexity and list notable papers that employ each algorithm to IDS. Hodo~\textit{et al.}~\cite{RefWorks:doc:5ae87684e4b0e00594a6a318} extend the ML discussion, the authors focus on the role that feature selection plays in the overall training and performance evaluation of ML techniques. An extensive discussion of features and how they impact the design and accuracy of IDS is plotted by Varma~\textit{et al.}~\cite{RefWorks:doc:5cdd83f9e4b09410f5d04260}. 
IDS characteristics are discussed by Debar~\textit{et al.}~\cite{RefWorks:doc:5cdd86aae4b0b127c2f6b596}, as well as, Amer and Hamilton~\cite{amer2010intrusion}. 

Hamed~\textit{et al.}~\cite{RefWorks:doc:5cdd85f5e4b02cf23c27b848} presents an overview of IDS components, listing them as (a)~pre-processing/feature extraction, (b)~pattern analyzer, which involves knowledge representation and learning processes, and finally (c)~decision making. They briefly discuss the benefits of each learning technique.

Additional IDS aspects are considered, for example, Amit~\textit{et al.}~\cite{RefWorks:doc:5ce279fae4b02cf23c28628b} list the various problems and challenges involved with using ML in building IDSs. 

The aforementioned manuscripts are further analyzed within Section~\ref{sec:recent-ids}.

Other perspectives included in recent studies focus on a single network architecture. For example, Ismail Butun~\textit{et al.}~\cite{RefWorks:doc:5ae87184e4b0a553e076cb0a} discusses Wireless Sensor Networks~(WSN), Chunjie Zhou~\textit{et al.}~\cite{RefWorks:doc:5ae87327e4b09318b71987f0} highlights IDS in industrial process automation while Ghaffarian and Shahriari~\cite{RefWorks:doc:5ae8746ae4b029aa3efa7c2b} study ML and Data Mining~(DM) techniques for software vulnerability. 

\vspace{2mm}
While these surveys provide valuable information on the design and the accuracy, none provide a detailed overview of the shortcomings of available datasets nor do they provide information on tools used to carry out attacks. In this manuscript, we address these shortcomings and provide detailed complementary work to build datasets that reflect current network threats. This manuscript is complementary to prior surveys by highlighting the shortcomings of current datasets and the claims of numerous studies on their abilities to detect deprecated attacks.
\section{IDS and Datasets Survey}
\label{sec:recent-ids}
In this section, prominent datasets are summarized, and their limitations are highlighted. Furthermore, recent IDSs are analyzed, discussing the algorithms used and the datasets the IDSs were evaluated against. Moreover, trends observed in the algorithms used by research over the past decade are discussed, highlighting a clear shift in the use of specific algorithms.

\subsection{Datasets}
\label{sec:dataset}
Researchers depended on benchmark datasets to evaluate their results. However, currently available datasets lack real-life characteristics of recent network traffic. This is  the reason that made most of the anomaly IDSs not applicable for production environments~\cite{RefWorks:doc:5ae872dee4b09318b71987e9}. Furthermore, IDS is unable to adapt to constant changes in networks (i.e. new nodes, changing traffic loads, changing topology, etc.).
Networks are constantly changing, for this reason depending solely on old datasets doesn't help the advancement of IDS. 
The process of generating new datasets should consider this constant change fact. For example, proposing a standard dataset generation platform with extendable functionality, would remove the burden of generating datasets from scratch and cope with concept drift in network patterns. This recommendation and others are further discussed in Section~\ref{sec:challenges}.

\begin{table*}[!ht]
    \centering
    \caption{Attacks Prominent Datasets}
    \begin{tabular}{|P{0.05\linewidth}|P{0.15\linewidth}|P{0.035\linewidth}|P{0.035\linewidth}|P{0.035\linewidth}|P{0.035\linewidth}|P{0.035\linewidth}|P{0.035\linewidth}|P{0.035\linewidth}|P{0.035\linewidth}|P{0.035\linewidth}|P{0.035\linewidth}|P{0.035\linewidth}|P{0.035\linewidth}|P{0.035\linewidth}|P{0.035\linewidth}|P{0.035\linewidth}|P{0.035\linewidth}|P{0.035\linewidth}|P{0.035\linewidth}|P{0.035\linewidth}|}
        \hline
         \rowcolor{gray!20}
         \multicolumn{21}{|c|}{\textbf{\textit{General Purpose Networks}}} \\
         \hline
         \rowcolor{gray!20}
          & & & & & & & & & \multicolumn{2}{c|}{Brute} & &  \multicolumn{5}{c|}{} & & & & \\
           \rowcolor{gray!20}
          & &  & & & & & & & \multicolumn{2}{c|}{Force} & &  \multicolumn{5}{c|}{Web} & & & & \\
          \cline{10-11} \cline{13-17} 
         \rowcolor{gray!20}
          Year & Dataset & \rotatebox{90}{Normal} & \rotatebox{90}{DoS} & \rotatebox{90}{DDoS} & \rotatebox{90}{Probe} & \rotatebox{90}{U2R} & \rotatebox{90}{R2L} & \rotatebox{90}{Infiltrating/Scanning} & \rotatebox{90}{SSH} & \rotatebox{90}{FTP} & \rotatebox{90}{Heartbleed} & \rotatebox{90}{Brute Force} & \rotatebox{90}{XSS} & \rotatebox{90}{Sql Injection} & \rotatebox{90}{Webshell} & \rotatebox{90}{DVWA} & \rotatebox{90}{Botnet} & \rotatebox{90}{Network and Host Events} & \rotatebox{90}{Port Scan} & \rotatebox{90}{Meterpreter} \\
         \hline
         2018 & CICIDS2018~\cite{sharafaldin2018toward} & \checkmark & \checkmark & \checkmark & - & - & - & \checkmark & \checkmark & - & \checkmark  & \checkmark & \checkmark & \checkmark & - & \checkmark & \checkmark & - & \checkmark & -  \\ \hline 
         2017 & CICIDS2017 \cite{RefWorks:doc:5b227bb8e4b07f83f15ddb45}  & \checkmark & \checkmark & \checkmark & - & - & - & \checkmark & \checkmark & \checkmark & \checkmark & \checkmark & \checkmark & \checkmark & - & -  & \checkmark & - & \checkmark & -\\ \hline

         2017 & CIC DoS dataset~\cite{RefWorks:doc:5b227c16e4b0dcaa93352175}  & \checkmark & \checkmark & - & - & - & - & - & - & - & - & - & - & - & - & - & - & - & - & - \\ \hline
         
         2017 \& 2013 & ADFA-IDS~\cite{RefWorks:doc:5cd94680e4b0661506bd7d82, RefWorks:doc:5b06d3c3e4b00f93d242857a}  & \checkmark & - & - & - & * & + & -  & \checkmark & \checkmark & - & - & - & - & \checkmark & - & - & - & - & \checkmark \\ \hline
     
         2017 & Unified Network Dataset~\cite{turcotte17}  & \checkmark & - & - & - & - & - & - & - & - & - & - & - & - & - & - & - & \checkmark & - & -\\ \hline
         
         2016 & DDoSTB~\cite{behal2016measuring} & \checkmark & - & \checkmark & - & - & - & - & - & - & - & - & - & - & - & - & - & - & - & - \\ \hline 
         
         2015 & Booters~\cite{RefWorks:doc:5cd970fde4b048754198ca22}  & - & - & \checkmark & - & - & - & - & - & - & - & - & - & - & - & - & - & - & - & - \\ \hline
         \multirow{3}{*}{2015} & TUIDS Coordinated Scan~\cite{bhuyan2015towards} & \checkmark & - & - & \checkmark  & - & - & \checkmark & - & - & - & - & - & - & - & - & - & -  & - & - \\ \cline{2-21} 

         & TUIDS DDoS~\cite{bhuyan2015towards} & \checkmark & - & \checkmark & -  & - & - & - & - & - & - & - & - & - & - & - & - & -  & - & - \\ \cline{2-21} 

         & TUIDS Intrusion~\cite{bhuyan2015towards} & \checkmark & \checkmark & - & \checkmark & - & - & - & - & - & - & - & - & - & - & - & - & - & - & - \\ \hline

        2014 & Botnet dataset~\cite{RefWorks:doc:5b227bf6e4b0d1cffc0657b8} & \checkmark & - & - & - & - & - & - & - & - & - & - & - & - & - & -  & \checkmark & - & - & -\\ \hline

          2012 & STA2018~\cite{amjad_al_tobi_2018}  & \checkmark & \checkmark & \checkmark & - & - & - & \checkmark & \checkmark & - & - & - & - & - & - & - & - & - & - & - \\ \hline

        2011 & CTU-13~\cite{garcia2014empirical}  & \checkmark & - & - & - & - & - & - & - & - & - & - & - & - & - & - & \checkmark & - & - & -\\ \hline
       
       2010 &   ISCXIDS2012 \cite{RefWorks:doc:5b06d318e4b0faf8604e40ce} & \checkmark & \checkmark & \checkmark & - & - & - & \checkmark & \checkmark & - & - & - & - & - & - & - & - & - & - & - \\ \hline
        
        2009 & Waikato~\cite{waikato} & - & - & \checkmark & - & - & - & - & - & - & - & - & - & - & - & - & - & - & - & -  \\ \hline
        2007 & CAIDA DDoS \cite{caida2007} & - & - & \checkmark & - & - & - & - & - & - & - & - & - & - & - & - & - & - & - & - \\ \hline
        
        1999 & NSL-KDD~\cite{RefWorks:doc:5b227bd4e4b0d1cffc0657b0} & \checkmark & \checkmark & - & \checkmark & \checkmark & \checkmark & - & - & - & - & - & - & - & - & - & - & - & - & - \\ \hline
      
        1999 & KDD'99~\cite{RefWorks:doc:5b06959ce4b078503013ca8c} &  \checkmark & \checkmark & - & \checkmark & \checkmark & \checkmark & - & - & - & - & - & - & - & - & - & - & - & - & - \\ \hline

        1998, 1999, 2000 &  DARPA~\cite{RefWorks:doc:5b0693fce4b0faf8604e2beb}  & \checkmark & \checkmark & - & \checkmark & \checkmark & \checkmark & - & - & - & - & - & - & - & - & - & - & - & - & - \\ \hline
        \hline
        \rowcolor{gray!20}
        \multicolumn{21}{|c|}{\textbf{\textit{Special Purpose Networks}}} \\
        \hline
        \rowcolor{gray!20}
        Year & Dataset & \multicolumn{5}{|c|}{IoT} &  \multicolumn{5}{|c|}{VPN} &  \multicolumn{5}{|c|}{Tor} & \multicolumn{4}{|c|}{SCADA} \\  
        \hline
       2018 & Bot-IoT~\cite{koroniotis2018towards} & \multicolumn{5}{|c|}{\checkmark} & \multicolumn{5}{|c|}{-} & \multicolumn{5}{|c|}{-} & \multicolumn{4}{|c|}{-} \\ \hline
        
        2017 & Anomalies Water System~\cite{RefWorks:doc:5ad0a812e4b0e18303dc2c3e, 10.1007/978-3-030-12786-2_1} & \multicolumn{5}{|c|}{-} & \multicolumn{5}{|c|}{-} & \multicolumn{5}{|c|}{-} & \multicolumn{4}{|c|}{\checkmark} \\  \hline
        
        2017 & IoT devices captures~\cite{miettinen2017iot} & \multicolumn{5}{|c|}{\checkmark} & \multicolumn{5}{|c|}{-} & \multicolumn{5}{|c|}{-} & \multicolumn{4}{|c|}{-} \\   \hline

        2016 & Tor-nonTor dataset~\cite{RefWorks:doc:5b22790ee4b032ce1ef1ee04} &  \multicolumn{5}{|c|}{-} & \multicolumn{5}{|c|}{-} & \multicolumn{5}{|c|}{\checkmark} & \multicolumn{4}{|c|}{-} \\  
         \hline
        2016 & VPN-nonVPN dataset~\cite{RefWorks:doc:5b22787ce4b034677a1ee449} & \multicolumn{5}{|c|}{-} &  \multicolumn{5}{|c|}{\checkmark} &  \multicolumn{5}{|c|}{-} & \multicolumn{4}{|c|}{-} \\  
        \hline 
        
        2015 & 4SICS ICS~\cite{SCADAICS9:online} & \multicolumn{5}{|c|}{-} & \multicolumn{5}{|c|}{-} & \multicolumn{5}{|c|}{-} & \multicolumn{4}{|c|}{\checkmark} \\  \hline
        \hline
        \rowcolor{gray!20}
        \multicolumn{21}{|c|}{\textbf{\textit{Mobile Applications}}} \\
        \hline
        \rowcolor{gray!20}
        Year & Dataset & \multicolumn{5}{|c|}{Benign} &  \multicolumn{5}{|c|}{Botnet} &  \multicolumn{5}{|c|}{Adware} & \multicolumn{4}{|c|}{Malware} \\  
        \hline
        2016 & Kharon Malware Dataset~\cite{kiss:hal-01300752} & \multicolumn{5}{|c|}{-} & \multicolumn{5}{|c|}{-} & \multicolumn{5}{|c|}{-} & \multicolumn{4}{|c|}{\checkmark}\\ \hline

        2015 - 2017 & Android Adware and General Malware Dataset~\cite{RefWorks:doc:5b227673e4b03aeb784ad16a} & \multicolumn{5}{|c|}{\checkmark} & \multicolumn{5}{|c|}{-} & \multicolumn{5}{|c|}{\checkmark} & \multicolumn{4}{|c|}{\checkmark}\\ \hline
        
        2010 - 2014 & Android Botnet dataset~\cite{RefWorks:doc:5b227757e4b0dcaa93351fbf} & \multicolumn{5}{|c|}{-} & \multicolumn{5}{|c|}{\checkmark} & \multicolumn{5}{|c|}{-} & \multicolumn{4}{|c|}{-}\\ \hline
        2010 - 2011 & Android Malware Genome~\cite{zhou2012dissecting} & \multicolumn{5}{|c|}{-} & \multicolumn{5}{|c|}{-} & \multicolumn{5}{|c|}{-} & \multicolumn{4}{|c|}{\checkmark}\\ \hline
        - & AndroMalShare~\cite{AndroMal22:online} & \multicolumn{5}{|c|}{-} & \multicolumn{5}{|c|}{-} & \multicolumn{5}{|c|}{-} & \multicolumn{4}{|c|}{\checkmark}\\ \hline
        \multicolumn{19}{c}{*: Adding new Superuser, +: C100}
    \end{tabular}
    \label{tab:cic-datasets}
\end{table*}

Datasets could either be real (i.e. recorded from a network set-up) or synthetic (i.e. simulated or injected traffic). Synthetic attack injection could be used to either introduce attacks to an existing dataset or balance the attack classes present in a dataset. Viegas~\textit{et al.}~\cite{RefWorks:doc:5ae872dee4b09318b71987e9} mentioned that for a dataset to be considered, it has to cover the following properties. (a)~Real network traffic (similar to production ones), (b)~valid, such that it has complete scenarios. (c)~Labeled, classifying each record as  normal or attack, (d)~variant, (e)~correct, (f)~can be updated easily. (g)~ Reproducible in order to give researchers space to compare across different datasets, and finally, (h)~shareable, hence it should not contain any confidential data. Additionally, Sharafaldin~\textit{et al.}~\cite{dataset-evaluationsharafaldin2018towards} mentions that (i)~having an appropriate documentation for the feature and dataset collection environment is an important aspect of IDS dataset.
Cordero~\textit{et al.}~\cite{RefWorks:doc:5cd4229ae4b004a2de355491} adds (j)~having high quality normal and (k)~excluding any disturbance or defects as further requirements for evaluation datasets.
Furthermore, for NIDS evaluation dataset, functional and non-functional requirements are elaborated in~\cite{RefWorks:doc:5cd44562e4b048754197304f}.

In this manuscript, we also identify two problems that impact research domains using datasets, whether they are synthetic or not. i) Sharing datasets is sometimes prohibited due to the data contained, hence, the research in the area is limited. ii) Simulating real-life scenarios and associated attacks is difficult due to the number of parameters required for the model to be viable. However, this manuscript provides a list of the most used and recent datasets.

\tablename~\ref{tab:cic-datasets} summarizes the available datasets and categorizes them based on the domain they belong to. Moreover, attacks found in each are presented. Extra remarks, including
the publication year, institute and attack classes details are
listed in \tablename~\ref{tab:attacks-remarks}. These datasets cover mobile applications, Virtual Private Networks (VPN), Tor Networks, IDS, Botnet, Network Flows and IoT. Some of the mentioned datasets are presented in~\cite{dataset-evaluationsharafaldin2018towards}. The evaluation includes DEFCON~\cite{RefWorks:doc:5b06cb93e4b0faf8604e3cb6}, CAIDA~\cite{RefWorks:doc:5b06cca2e4b053be15af02aa}, LBNL~\cite{RefWorks:doc:5b06cec6e4b0afb29cb50b91}, CDX~\cite{RefWorks:doc:5b06cfc8e4b00f93d24283ae}, Kyoto~\cite{RefWorks:doc:5b06d0e8e4b0fb7e4ea4aebc}, Twente~\cite{RefWorks:doc:5b06d152e4b0fb7e4ea4aef3}, UMASS~\cite{RefWorks:doc:5b06d226e4b0fb7e4ea4afa8} and ADFA~\cite{RefWorks:doc:5b06d3c3e4b00f93d242857a}. 

Ring~\textit{et  al.}~\cite{RefWorks:doc:5cd94fbfe4b0de98174c4e1c} comprehensively overview of NIDS datasets covering their main features, data format, anonymity, size, availability, recording environment, balancing, etc\ldots The authors list the datasets and their corresponding values in each of the aforementioned criteria, leaving the choice for researchers to make based on their use-case and scenario. 
On the contrary, Gharib~\textit{et  al.}~\cite{7885840} propose datasets score based on the attacks' coverage, protocols' coverage, metadata availability, anonymity, heterogeneity and labeling. 
While the authors evaluate attacks in the datasets and present a scientific comparison, the authors fail to provide a detailed analysis of the broader impact of their analysis.

Furthermore, due to the sparsity of the details supplementing the available datasets, the task of evaluating and ranking datasets would introduce unfair results. For example, a dataset that realistically represents background and attack traffic is better than a dataset that doesn't. However, there is no standard metric to evaluate how realistic the generation is, as well as, this information is not released with the dataset.

\subsection{IDS and Associated Datasets Analysis}
\label{subsec:recent-ids}
In this section, a survey of recent ML IDS is provided, analyzing the associated datasets, and their shortcomings.
IEEE Xplore and Google Scholar queries were made using "Intrusion Detection System*'' OR ``IDS*'' filtering the dates to include manuscripts published in the last decade. The filtration was made to have a  wide coverage of datasets, ML techniques and detected attacks. A total of \textbf{85} published manuscripts in the period of $[2008-2020]$ are analyzed. Analysis of older IDS ML techniques and used features for the period 2004 - 2007 was previously conducted by Nguyen and Armitage~\cite{4738466}. They discuss the limitations of port-based and payload-based classification and the emerging use of ML techniques to classify IP traffic.

\tablename~\ref{tab:appendix} summarizes the pre-eminent (i.e., most cited) IDS research from the past decade. Each IDS is mentioned with a list of the  algorithms used and the datasets that the IDS was evaluated against. Moreover, the  attacks detected are also listed. 
The algorithmic trends are then discussed alongside the attacks included in the datasets used.

\begin{figure}[bth]
    \centering\includegraphics[width=\linewidth, keepaspectratio=true] 
        {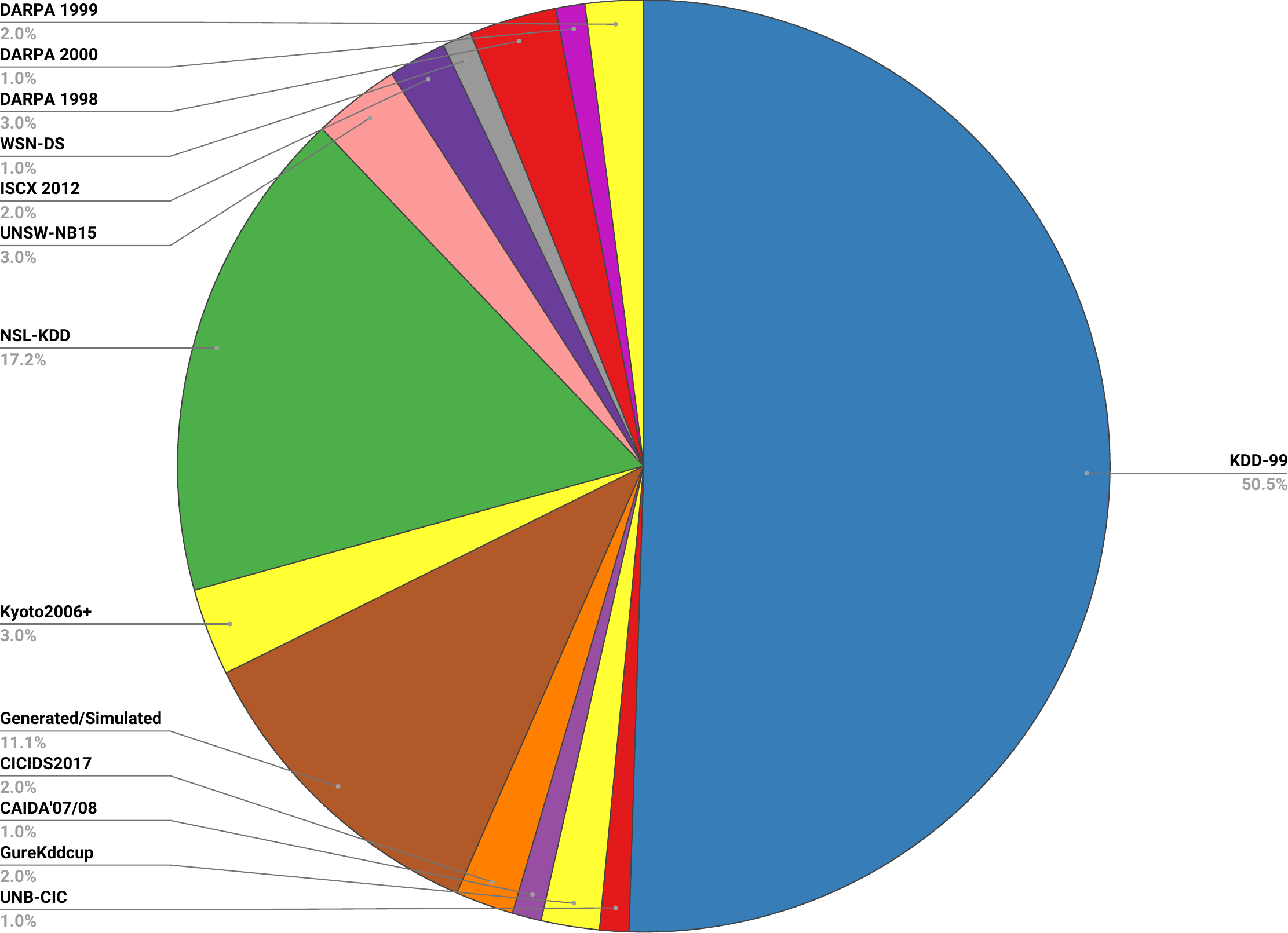}
    \caption{Datasets used for IDS Evaluation Distribution (85 IDSs Manuscripts listed in Table~\ref{tab:appendix})}
      \label{ids_dataset_dist}
\end{figure}

\figurename~\ref{ids_dataset_dist} shows the distribution of datasets used for research in the last decade. Only 11\% of the mentioned IDSs used generated or simulated datasets. It is also clear through this analysis that most datasets lack real-life properties, which were previously mentioned in Section~\ref{sec:dataset}. \figurename~\ref{ids_dataset_dist} also highlights the use of KDD-99 as the dataset of choice. 
Amjad Al Tobi and Ishbel Duncan~\cite{RefWorks:doc:5bbe0843e4b0e904d9061fbb} provide a comprehensive analysis of the drawbacks of the KDD'99 dataset. Moreover, Siddique~\textit{et al.}~\cite{siddique2019kdd} provide a timeline for KDD datasets family. The provided timeline  show both the different criticism points and the UCI Lab warning not to use KDD Cup'99 dataset, which further emphasizes the drawbacks of using KDD Cup'99 in the current IDS research. The second most used dataset is the DARPA datasets. DARPA datasets fail to accurately represent current attacks due to their age. Moreover, the use of the KDD'99 and DARPA datasets lead to an endemic situation, numerous results reported in literature claim detection results which are not applicable in real-world scenarios. The shortcomings of the DARPA dataset are analyzed by M. Mahoney and P. Chan~\cite{RefWorks:doc:5bbe0538e4b08bd7cfb5b064} and John McHugh~\cite{McHugh:2000:TID:382912.382923}.
Alongside the limitations of each dataset, they are also deprecated, hence,  demonstrating the inability of the IDSs presented in~\tablename~\ref{tab:appendix} to cope with the most recent attacks and threats.

\begin{figure}[bth]
    \centering\includegraphics[width=\linewidth, keepaspectratio=true]{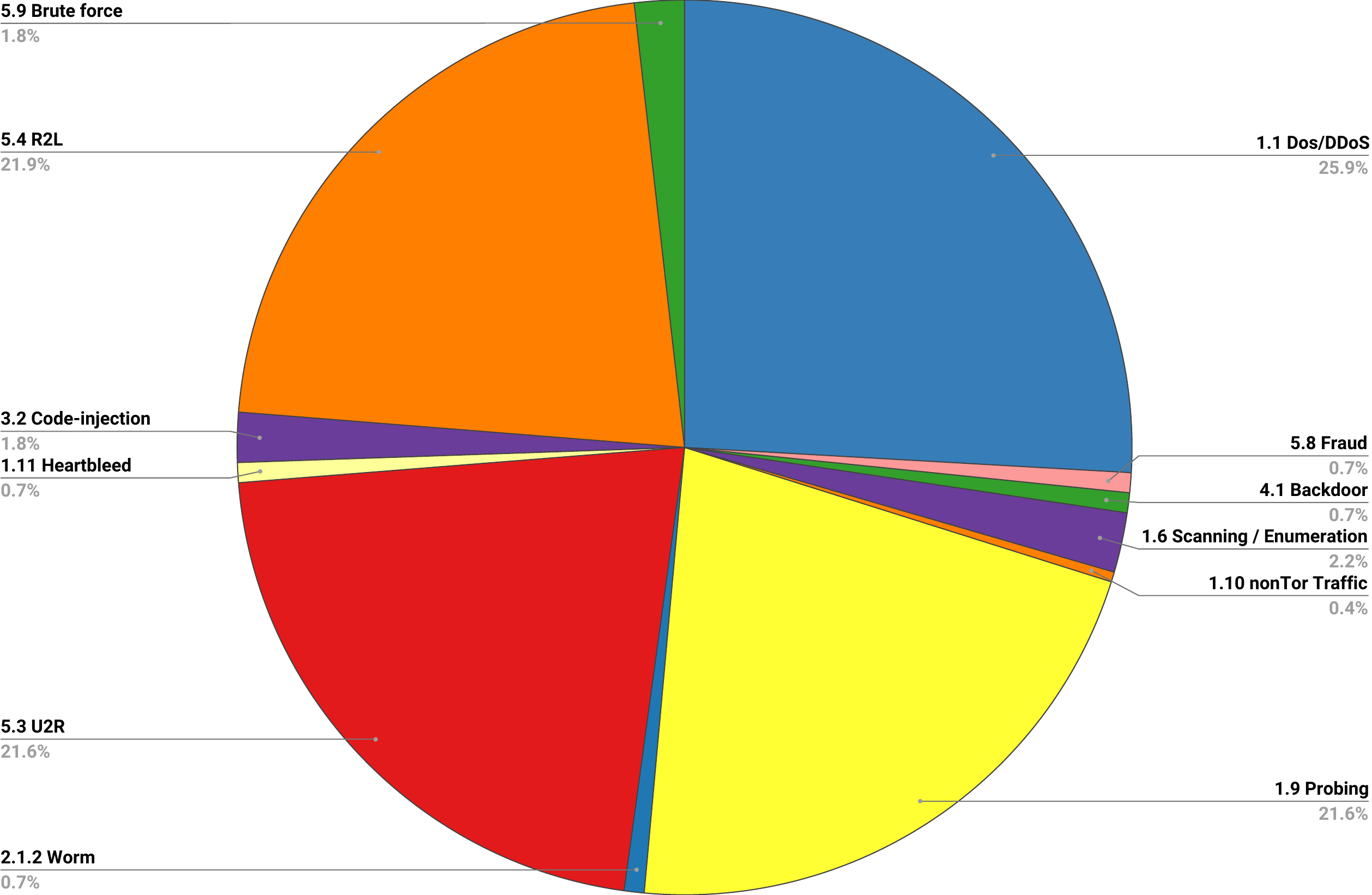}
    \caption{Covered Attacks in Discussed IDS (85 IDSs Manuscripts listed in Table~\ref{tab:appendix})}
    \label{ids_attacks_dist}
\end{figure} 

\figurename~\ref{ids_attacks_dist} visualizes the attacks detected by the different IDSs presented in \tablename~\ref{tab:appendix}. It is shown that the four attacks available in the KDD-99 dataset are the most covered, namely; DoS/DDoS, Probing, R2L, U2R. Moreover, only 12 attacks are listed in \figurename~\ref{ids_attacks_dist} which highlights potential limitations of these IDS to cope with the broad range of attacks and zero-day attacks. To tackle the detection of zero-day attacks, there is a need to build extendable datasets that could be used to train machine learning models used for anomaly detection. By employing extendable datasets and a standardized method for dataset generation, alongside the advancement in ML~\cite{Parrend2018, RefWorks:doc:5cdea3dee4b09ae3c8870463}, zero-day detection could be integrated into anomaly-based IDSs.
Later in Section~\ref{sec:Threats}, our presented threat taxonomy highlights the percentage of attack coverage achieved by current IDSs.

To further analyze the last decade research on IDSs, it is important to consider the algorithms used. Anomaly-based IDSs are based on identifying patterns that define normal and abnormal traffic. These IDSs can be classified into subcategories based on the training method used as aforementioned in Section~\ref{sec:IDS}.
These categories are identified respectively as statistical, knowledge-based and ML based. Statistical includes univariate, multivariate and time series. Knowledge-based uses finite state machines and rules like case-based, n-based, expert systems and descriptor languages. ML algorithms include artificial neural networks, clustering, genetic algorithms, Deep Learning~(DL), etc.

\begin{figure}
    \centering
    \begin{subfigure}{\linewidth}
    \centering
    \includegraphics[width=1\linewidth, trim={0cm 0.1cm 0cm 0.1cm}, clip]{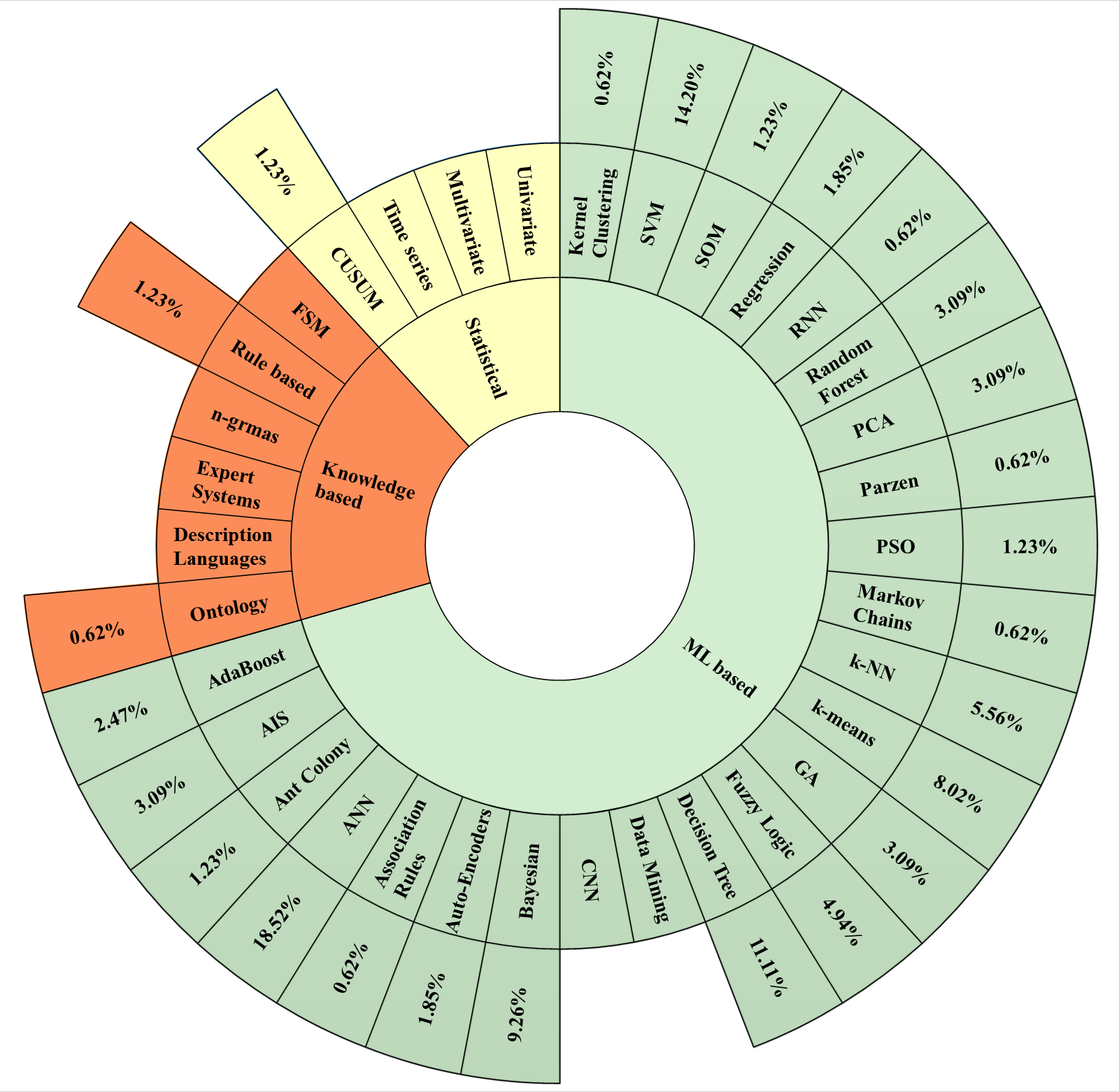}
    \caption{Distribution of all algorithms categories}
    \vspace{2mm}
    \end{subfigure}
     \begin{subfigure}{\linewidth}
    \centering
    \includegraphics[height=0.9\linewidth, angle=90]{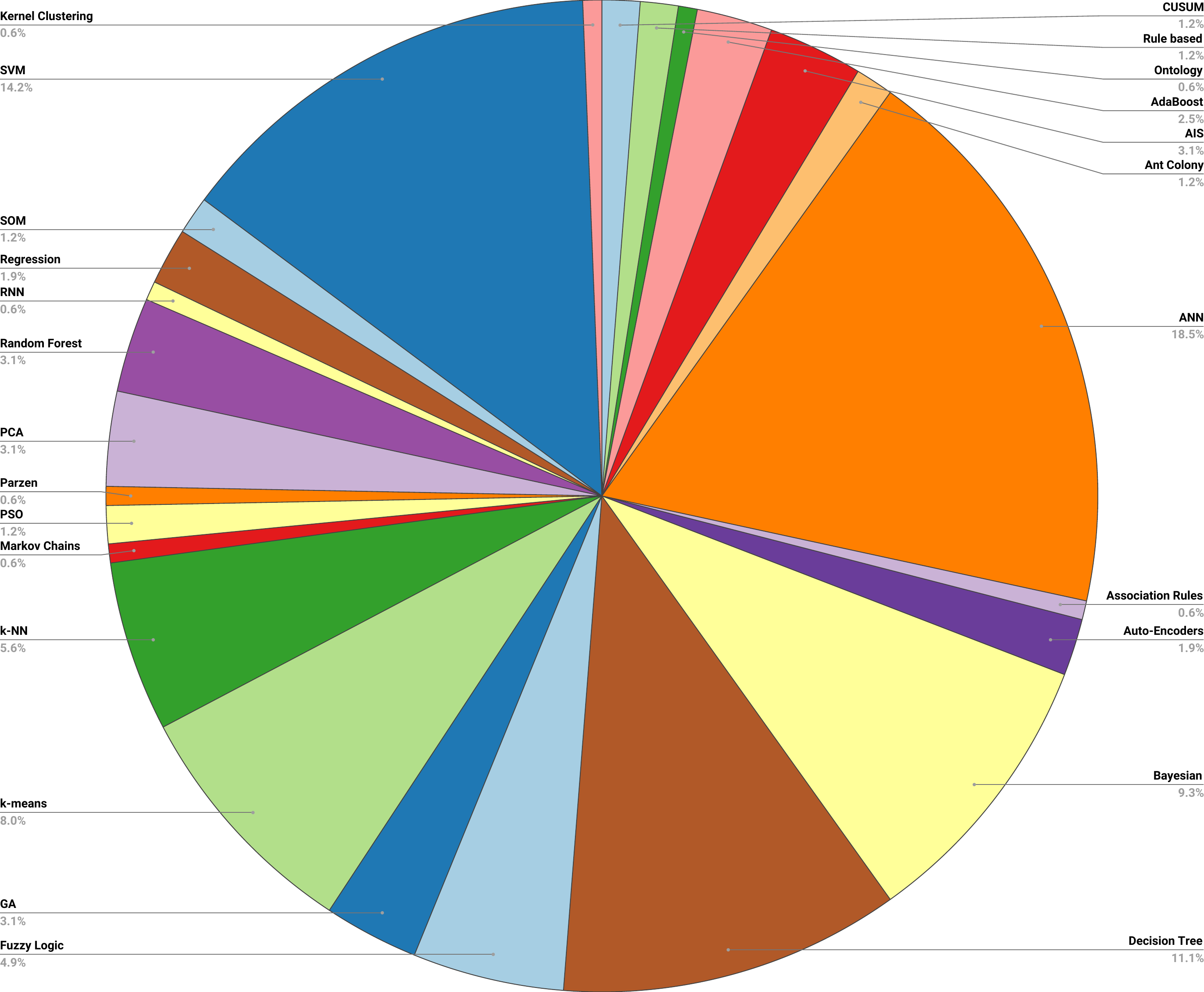}
    \caption{Distribution of used algorithms discussed in \tablename~\ref{tab:appendix}}
    \end{subfigure}
    \caption{Algorithms Usage Distribution in the discussed IDS (85 IDSs Manuscripts listed in  Table~\ref{tab:appendix})
    \\
    {\footnotesize Such that: 
    \quad AdaBoost: Adaptive Boosting \\ AIS: Artificial Immune System \quad 
    ANN: Artificial Neural Network \\  CNN: Convolutional Neural Network \quad \hspace{2ex}  CUSUM: Cumulative Sum \\
   FSM: Finite State Machine \quad \hspace{11ex} GA: Genetic Algorithms \\ k-NN: k-Nearest Neighbors \quad \hspace{10ex}ML: Machine Learning \\  PCA: Principal Component Analysis \quad \hspace{1ex}
   PSO: Particle Swarm Optimization \\  RNN: Recurrent Neural Network \quad \hspace{6ex} SOM: Self-Organizing Map \\ SVM: Support Vector Machine}
    }
    \label{ids_algo_dist}
\end{figure}

\figurename~\ref{ids_algo_dist}~(a) highlights the dominance of ML algorithms employed when building an IDS. As shown, both statistical and knowledge-based algorithms are less represented. This dominance is due to the significant use of ML techniques in various research domains. \figurename~\ref{ids_algo_dist}~(a)
is organized by categories~(Inner Circle), subcategories~(Centre Circle) and finally, the percentage of the IDSs presented in \tablename~\ref{tab:appendix} using these algorithms~(Outer Circle). 
\figurename~\ref{ids_algo_dist}~(b) on the other hand, provides a visualization of  the distribution of the algorithms used by the IDSs presented in \tablename~\ref{tab:appendix}. The dominance of ANN, SVM and k-means as the most used algorithms is reasoned by their ability to discriminate between benign and attack classes given a feature set. However, it is important to mention that leveraging new ML techniques and adapting ones from other domains will advance the development of the next decade's IDSs. 

\section{Threats Taxonomy}
\label{sec:Threats}
Building a generic and modular taxonomy for security threats is of high importance in order to help researchers and cyber-security practitioners build tools capable of detecting various attacks ranging from known to zero-day attacks.

Kendall~\textit{et al.}~\cite{RefWorks:doc:5ae86ac8e4b0e00594a69ff6} proposes one of the earliest classifications of intrusions~\cite{RefWorks:doc:5ae872dee4b09318b71987e9}. Kendall classifies intrusions into four categories namely: Denial of Service (DoS), Remote to Local (R2L), User to Root (U2R) and Probing. In DoS, the attacker tends to prevent users from accessing a given service. When the attacker tried to gain authorized access to the target system, either by gaining local access or promoting the user to a root user, these attacks were classified as R2L and U2R respectively. Finally, probing was defined as an attacker actively footprinting a system for vulnerabilities.  

Donald Welch classifies common threats in wireless networks into seven attack techniques (Traffic Analysis, Passive Eavesdropping, Active Eavesdropping, Unauthorized Access, Man-in-the-middle, Session Hijacking and Replay) \cite{welch2003wireless}. In a paper by Sachin Babar~\textit{et al.}~\cite{babar2010proposed}, the problem is addressed from a different perspective. Threats are classified according to the Internet of things security requirements (identification, communication, physical threat, embedded security and storage management). Specific domain taxonomies have also grabbed the attention of researchers. David Kotz~\cite{kotz2011threat} discusses privacy threats in the mobile health (mHealth) domain. In the same manner, Keshnee Padayachee~\cite{padayachee2012taxonomy}, shows the security threats targeting compliant information and Monjur Ahmed and Alan T. Litchfield~\cite{ahmed2018taxonomy} works on threats from a cloud computing point of view. 

This section classifies network threats based on the layers of the OSI model, provides examples of attacks for different threat types and presents a taxonomy associating network threats and the tools used to carry out attacks.
The taxonomies aim at helping researchers building IDSs, but more importantly by associating the threats to the OSI model and benchmarking the threats to the tools used to carry attack or take advantage of specific vulnerabilities, the taxonomies aim to achieve higher accuracies and reduce the number of false positives of current IDS~\cite{RefWorks:doc:5ae339c4e4b0e00594a5c5cc} and build better datasets.

\subsection{Threat Sources}
\label{sources-threats}

\begin{figure*}[!ht] 
    \includegraphics[angle=90, height=0.95\textheight, keepaspectratio=true] 
        {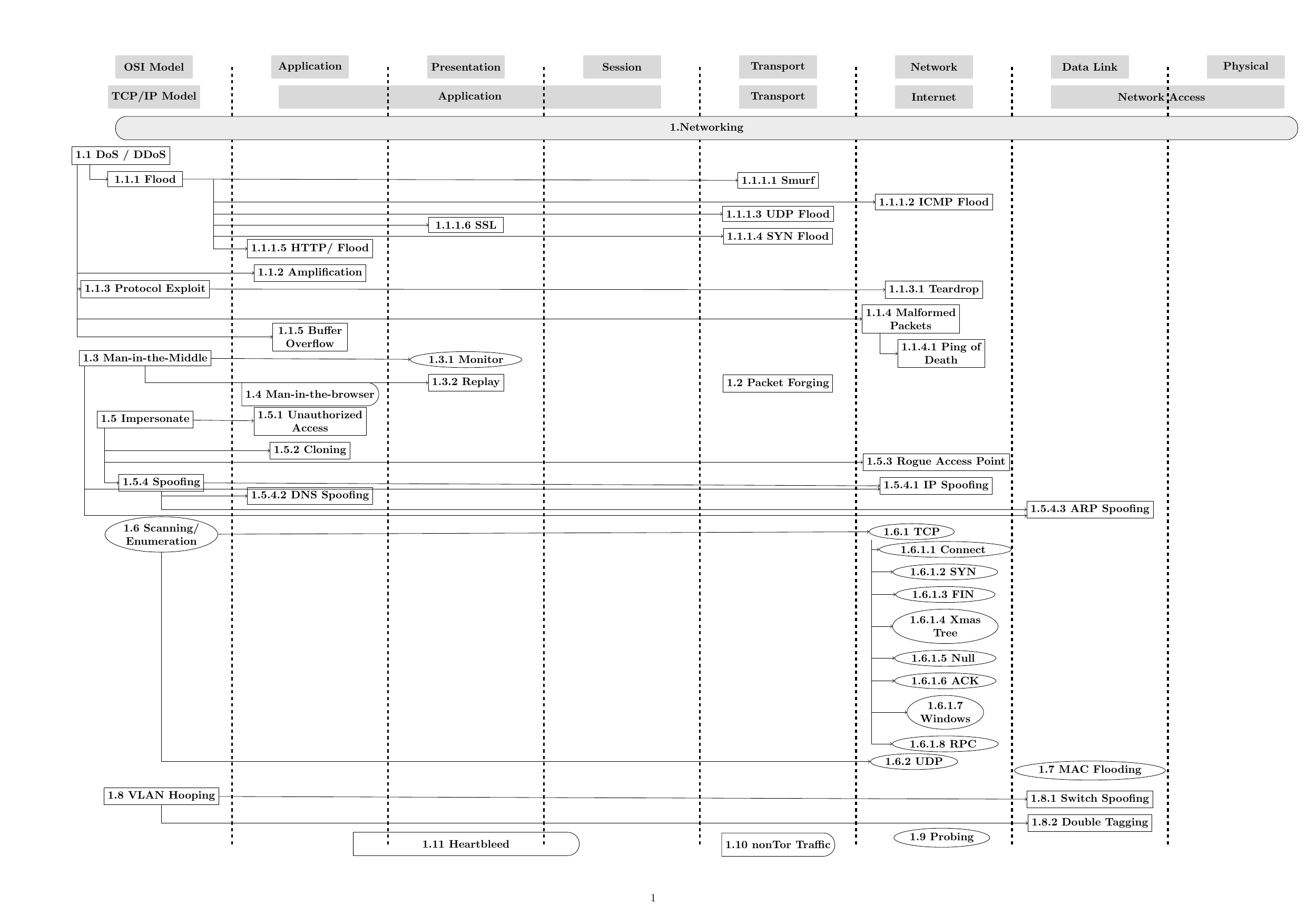}
     \caption{Taxonomy of threats (1 of 3)}
\end{figure*}

\begin{figure*}[!ht]\ContinuedFloat
 \includegraphics[angle=90, height=0.95\textheight, keepaspectratio=true] 
        {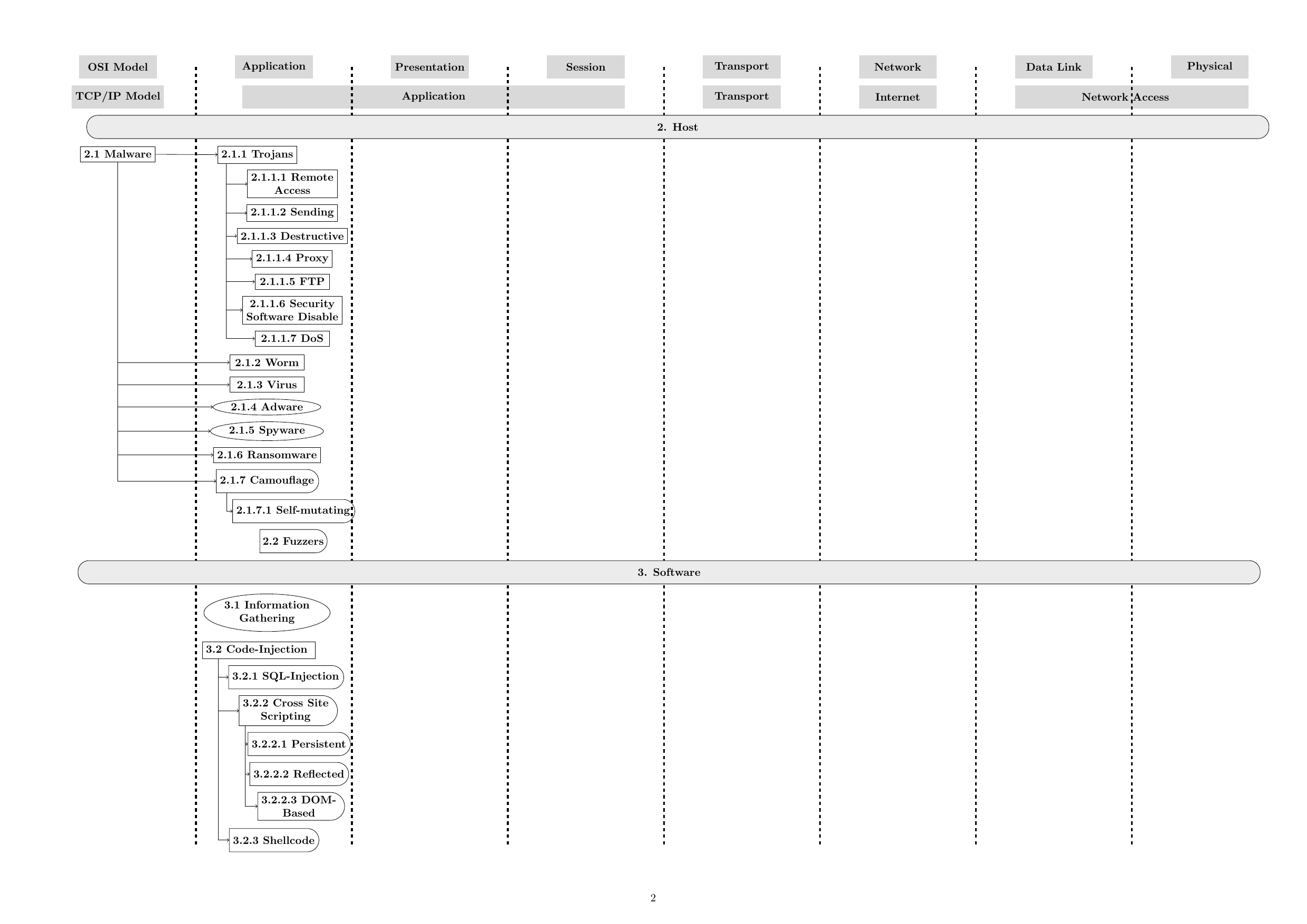}
 	\caption{Taxonomy of threats (2 of 3)}
\end{figure*}

\begin{figure*}[!ht]\ContinuedFloat
 \includegraphics[angle=90, height=0.953\textheight, keepaspectratio=true] 
        {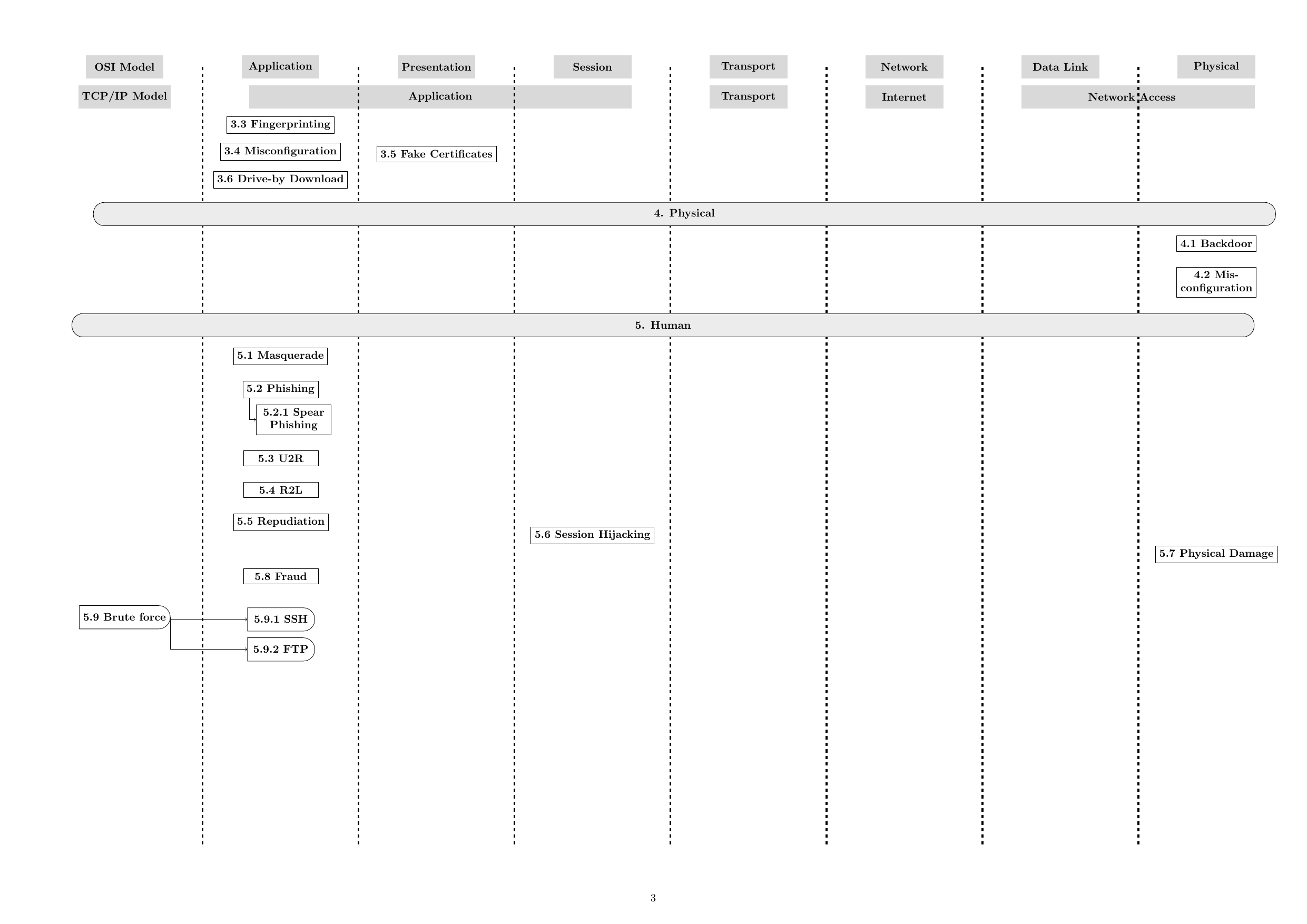}
 	\caption{Taxonomy of threats (3 of 3)}
  	\label{taxonomy_1}
\end{figure*}

\figurename~\ref{taxonomy_1} identifies network threats and provides a  classification according to the following criteria; (I) source of the threat, (II) affected layer based on Open Systems Interconnection (OSI) model and (III) active and passive threats. The different threats are described hereafter.

As shown, attacks can target a single layer of the OSI model, but it is important to highlight that other layers may also be affected. The taxonomy presented in this manuscript focuses on the main target layer of attack. An attack is also described to be active if it affects information, performance, or any aspect of the media on which it is running. In contrast to active attacks, during passive attacks the attacker is concerned with either gathering information or monitoring the network. These can be identified by their shape in \figurename~\ref{taxonomy_1}. Active attacks are represented by a \textit{rectangle shape}, whilst passive attacks are represented by an \textit{oval shape}. Attacks like adware~(\figurename~\ref{taxonomy_1} - 2.1.3), spyware~(\figurename~\ref{taxonomy_1} - 2.1.4) and information gathering~(\figurename~\ref{taxonomy_1} - 3.1) are considered passive attacks. DoS~(\figurename~\ref{taxonomy_1} - 1.1), Impersonation~(\figurename~\ref{taxonomy_1} - 1.4) and Virus~(\figurename~\ref{taxonomy_1} - 2.1.2) are forms of active attacks. However, some attacks cannot be considered active or passive until their usage is known. An example of this case is SQL-injection, if it is used for querying data from a database then it is passive. However, if it is used to alter data, drop tables or relations then the attack can be considered as active. 

\subsubsection{Network Threats}
Threats are initiated based on a flow of packets sent over a network. Two of the most common forms of network threats are Denial of Service~(DoS) and Distributed Denial of Service~(DDoS)~(\figurename~\ref{taxonomy_1} - 1.1), where an attacker floods the network with requests rendering the service unresponsive. During these attacks, legitimate users cannot access the services. Note that common anomalies known as `Flash Crowds' are often mistaken with DoS and DDoS attacks~\cite{jung2002flashcrowds}. Flash Crowds happen when a high flow of traffic for a certain service or website occurs. This happens immediately upon the occurrence of a significant event. For example, breaking news, sales events, etc. DoS and DDoS can be divided into four categories including flood attacks~(\figurename~\ref{taxonomy_1} - 1.1.1), amplification attacks~(\figurename~\ref{taxonomy_1} - 1.1.2), protocol exploit~(\figurename~\ref{taxonomy_1} - 1.1.3), and malformed packets~(\figurename~\ref{taxonomy_1} - 1.1.4). These are defined respectively  through attack examples. Smurf attacks~(\figurename~\ref{taxonomy_1} - 1.1.1.1) depends on generating a large amount of ping requests. Overflows~(\figurename~\ref{taxonomy_1} - 1.1.1.2) occurs when a program writes more bytes than allowed. This occurs when an attacker sends packets larger than 65536 bytes (allowed in the IP protocol) and the stack does not have an appropriate input sanitation in place. The ping of Death~(\figurename~\ref{taxonomy_1} - 1.1.4.1) attack occurs when  packets are too large for routers and splitting is required. The Teardrop~(\figurename~\ref{taxonomy_1} - 1.1.3.1) attack takes place when an incorrect offset is set by the attacker. Finally, the SYN flood~(\figurename~\ref{taxonomy_1} - 1.1.1.3) attack happens when the host allocates memory for a huge number of TCP SYN packets. 

Packet forging (\figurename~\ref{taxonomy_1} - 1.2) is another form of networking attack. Packet forging or injection is the action where the attacker generates packets that look the same as normal network traffic. These packets can be used to perform unauthorized actions and steal sensitive data like: login credentials, personal data, credit card details, Social Security Numbers~(SSN) numbers, etc. When the attacker passively monitors or intercepts communications between two or more entities and starts to control the communication, this attack is referred to as a 'Man in the Middle' attack~(\figurename~\ref{taxonomy_1} - 1.3). Unlike `Man in the Middle' attack, a `Man In The Browser' attack~(1.4) intercepts the browser to alter or add fields to a web page asking the user to enter confidential data.
Impersonation~(\figurename~\ref{taxonomy_1} - 1.5) or pretending to be another user can take different forms. The attacker may impersonate a user to gain higher security level and gain access to unauthorized data~(\figurename~\ref{taxonomy_1} - 1.5.1) or use cloning~(\figurename~\ref{taxonomy_1} - 1.5.2). Cloning is a common attack in social networks to impersonate an individual to leverage information. Rogue access points~(\figurename~\ref{taxonomy_1} - 1.5.3) are other impersonation forms in wireless networks. During an IP spoofing attack~(\figurename~\ref{taxonomy_1} - 1.5.4.1) an attacker spoofs an IP address and sends packets impersonating a legitimate host. DNS spoofing - also known as DNS cache poisoning ~(\figurename~\ref{taxonomy_1} - 1.5.4.2) is another type of spoofing. The attacker redirects packets by poisoning the DNS. Finally, ARP spoofing~(\figurename~\ref{taxonomy_1} - 1.5.4.3) is used to perform attacks like Man In the Middle, in order to dissociate legitimate IP and MAC addresses in the ARP tables of victims. 

Scanning/enumeration are an essential step for initiating attacks. During scanning~(\figurename~\ref{taxonomy_1} - 1.6), the attacker starts with searching the network for information such as: active nodes, running operating systems, software versions, etc. As defined in~\cite{hacking_book}, scanning has many forms, using protocols such as TCP~(\figurename~\ref{taxonomy_1} - 1.6.1) or UDP~(\figurename~\ref{taxonomy_1} - 1.6.2). The last two examples of network attacks are media access control~(MAC) address flooding~(\figurename~\ref{taxonomy_1} - 1.7), and VLAN hopping attack~(\figurename~\ref{taxonomy_1} - 1.8). In MAC flooding~(\figurename~\ref{taxonomy_1} - 1.7), the attacker is targeting the network switches and as a result, packets are redirected to the wrong physical ports, while the VLAN hopping attack has two forms of either switch spoofing~(\figurename~\ref{taxonomy_1} - 1.8.1) or double tagging~(\figurename~\ref{taxonomy_1} - 1.8.2).

\subsubsection{Host Threats}
Host attacks target specific hosts or systems by running malicious software to compromise or corrupt system functionalities. Most host attacks are categorized under the malware~(\figurename~\ref{taxonomy_1} - 2.1) category. This includes worms, viruses, adware, spyware, Trojans and ransomware.
Viruses are known to affect programs and files when shared with other users on the network, whilst worms are known to self-replicate and affect multiple systems. Adware is known for showing advertisements to users when surfing the Internet or installing software. Although adware is less likely to run malicious code, it can compromise the performance of a system. Spyware gathers information such as documents, user cookies, browsing history, emails, etc. or monitors and tracks user actions. 
Trojans often look like trusted applications, but allow an attacker to control a device. 
Furthermore, camouflage malware~(\figurename~\ref{taxonomy_1} - 2.1.7) evolved over time reaching polymorphic and metamorphic techniques in 1990 and 1998 respectively~\cite{rad2012camouflage, Galal2016}. For example, self-mutating malware could use numerous techniques, such as, instruction substitution or permutation, garbage insertion, variable substitutions and control-flow alteration~\cite{4140990}.
Last, ransomware is a relatively new type of malware where the system is kept under the control of the attacker - or a third entity - by encrypting files until the user/organization pays a ransom~\cite{RefWorks:doc:5ae8785ae4b0c16216f56d11}.

\subsubsection{Software Threats}
Code injection~(\figurename~\ref{taxonomy_1} - 3.2) can include SQL Injection to query a database, resulting in obtaining confidential data, or deleting data by dropping columns, rows or tables. Cross-site scripting (XSS) is used to run malicious code to steal cookies or credentials. XSS has three main categories. The first is persistent/stored XSS~(\figurename~\ref{taxonomy_1} - 3.2.2.1), in this case, a script is saved to a database and is executed every time the page is loaded. The second is Reflected XSS~(\figurename~\ref{taxonomy_1} - 3.2.2.2), where the script is part of a HTTP request sent to the server. The last is DOM-based XSS (\figurename~\ref{taxonomy_1} - 3.2.2.3) which can be considered as an advanced type of XSS. The attacker changes values in the Document Object Model~(DOM) e.g. document location, document URL, etc. DOM-based XSS is difficult to detect as the script is never transferred to the server. Drive-by or download~(\figurename~\ref{taxonomy_1}-3.6) is another software threat that requires no action from the user, however, the malicious code is automatically downloaded. It contributed to 48\% of all web-based attacks in 2017~\cite{RefWorks:doc:5cdd40e1e4b029a7217e42bf,neely2017threat} and is considered one of the main threats in 2019~\cite{RefWorks:doc:5cdd415ee4b0d691aab98edf}.
Fingerprinting~(\figurename~\ref{taxonomy_1} - 3.3) and misconfiguration are also forms of software threats. Fake server certificates~(\figurename~\ref{taxonomy_1} - 3.5) are considered alarming and should be considered while analyzing communications as they could deceive the browser/user thinking that the connection is secure. This could result in phishing websites looking legitimate. Moreover, they could be used as a seed to perform other attacks like Man-in-the-Middle.

\begin{figure*}[!th]
   \centering
   \includegraphics[width=0.6\textwidth]{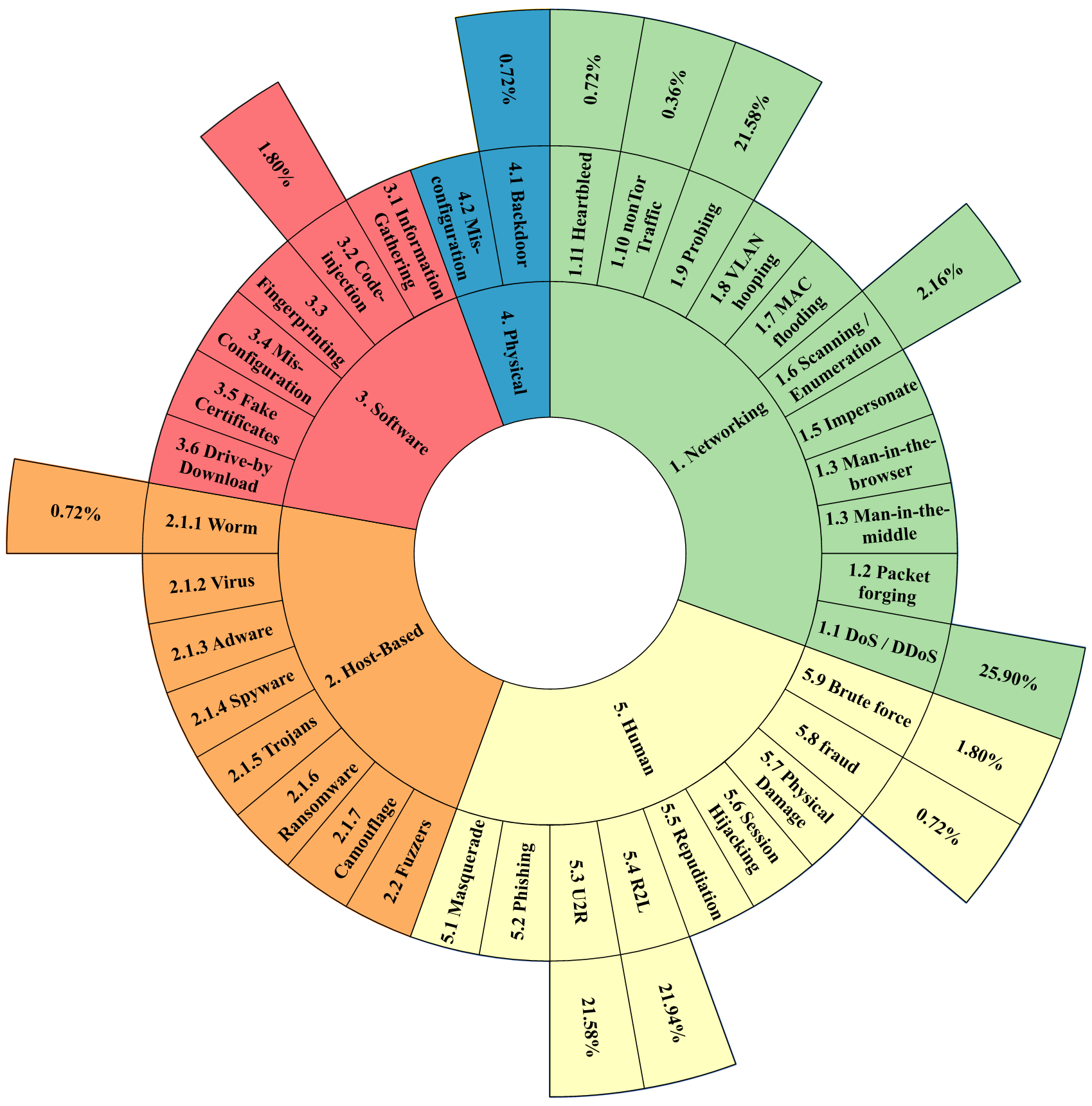}
   \caption{Distribution of Covered Attacks in Discussed IDS (85 IDSs Manuscripts listed in Table~\ref{tab:appendix})}
   \label{fig:ids-attacks-2}
\end{figure*}

\subsubsection{Physical Threats}
Physical attacks are a result of a tempering attempt on the network hardware (edge, or other devices) or its configuration. This can include changing configurations~(\figurename~\ref{taxonomy_1} - 4.2) and introducing backdoors (i.e. The Evil Maid).

\subsubsection{Human Threats}
The last category of networking attacks is one based on human actions. These include user masquerade~(\figurename~\ref{taxonomy_1} - 5.1). Phishing is another form of human attacks where the attacker uses emails or other electronic messaging services to obtain credentials or confidential data. When a user attempts to obtain higher privileges, it is considered a human attack like User to Root~(\figurename~\ref{taxonomy_1} - 5.3) and Remote to Local R2L~(\figurename~\ref{taxonomy_1} - 5.4). Additionally, a user can be denied an action such as repudiation attack~(\figurename~\ref{taxonomy_1} - 5.5). Human attacks can also include session hijacking or sniffing, these attacks are based on the attacker gaining access over an active session to access cookies and tokens. 

\vspace{1mm} 
\noindent
Based on the taxonomy discussed in \figurename~\ref{taxonomy_1} and the recent IDSs discussed in Section~\ref{subsec:recent-ids}, it can be seen that there are many threats that are not addressed by recent IDSs. \figurename~\ref{fig:ids-attacks-2} visualizes all the threats mentioned in the taxonomy. The associated percentage represents attacks covered by the IDSs discussed in \tablename~\ref{tab:appendix}. As shown a large number of attacks~(72\%) are not covered. Hence, the network threat taxonomy aims at addressing the following:
\begin{itemize}
    \item Help researchers generate datasets that cover non-addressed attacks.
    \item Provide an up-to-date taxonomy of attacks allowing to  measure threats covered by datasets and the ability of IDSs to detect these threats
    \item Provide a structured way to address and represent threats and attacks.
\end{itemize}

\begin{figure*}[!th]
\centering
\includegraphics[height=0.95\textheight,keepaspectratio=true, trim={1cm 5cm 1cm 1cm},clip]{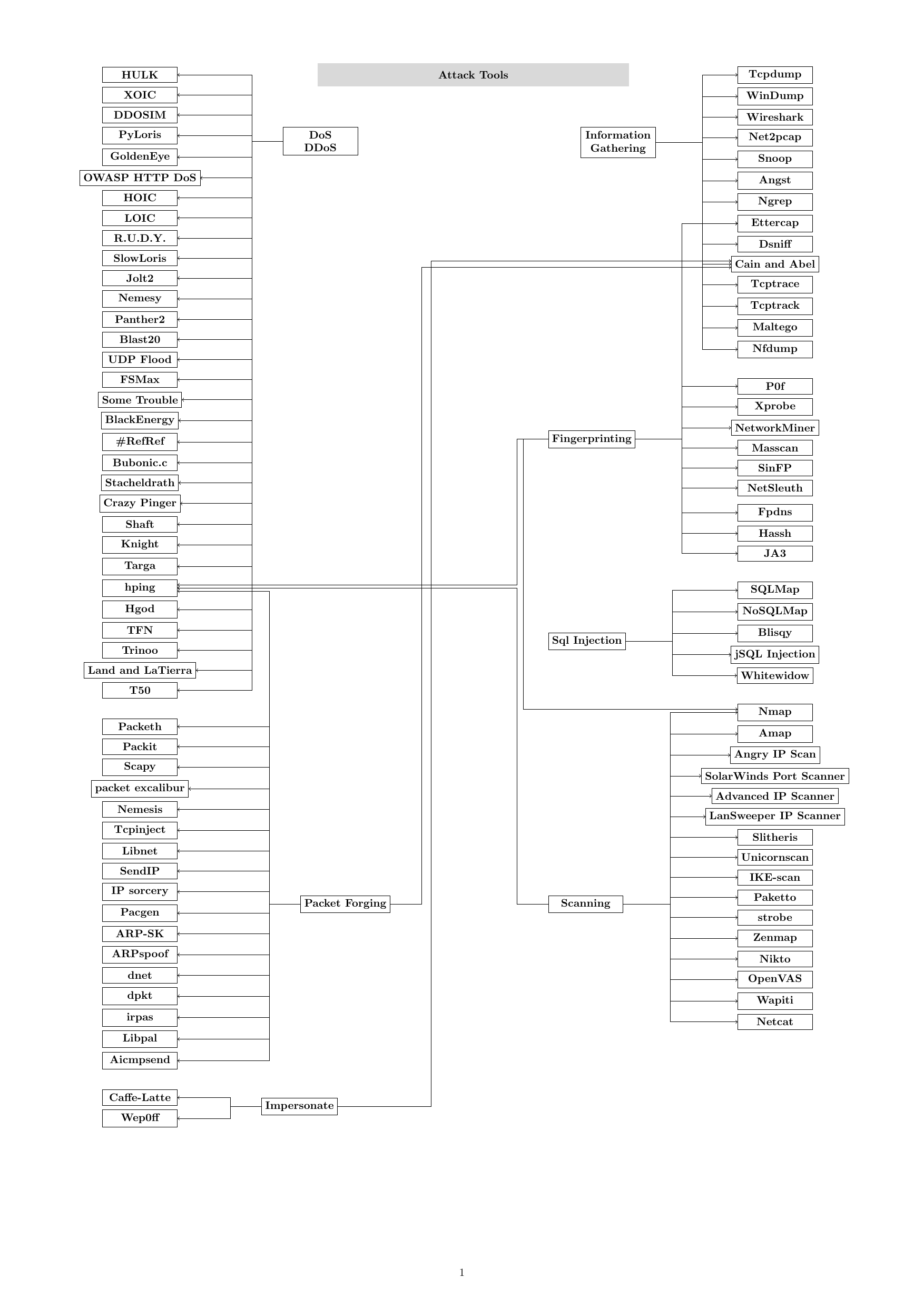}
     \caption{Attacking tools}
      \label{tools-examples}
\end{figure*}

\subsection{Attacking Tools}
Many tools~\cite{hacking_book}~\cite{RefWorks:doc:5ae86fb0e4b0c16216f56b55} have been developed to initiate different attacks. \figurename~\ref{tools-examples} shows the main tools classified by the attacks they are used for.
This can be used by researchers when building an IDS for a specific threat, then the associated tools are ones of interest. For example, for an IDS classifying impersonation attacks, Caffe-Latte, Hirte, EvilTwin and Cain and Abel are tools to check. Yaga and SQL attacks are tools used for U2R and so on.
\section{Challenges and Recommendations}
In this section, our findings are outlined based on the discussion in Section~\ref{sec:recent-ids} and Section~\ref{sec:Threats}. A list of limitations is reviewed then the recommendations are listed.

\label{sec:challenges}
\subsection{Limitations and Challenges}
The limitations and challenges in datasets used in IDSs can be summarized in the following:
\begin{itemize}
\item \textbf{Attacks Coverage}: As shown in this work only~33.3\% of known attacks are covered in publicly available datasets reviewed. This is considered one of the biggest challenges preventing IDSs to be used in real-life environments. 

\item \textbf{Real-life Simulation}: Only~11\% of the past decade IDSs use recent and/or real-life generated or simulated datasets. This demonstrates a flaw in the development of IDSs but highlights their limited ability to cope with the emerging needs. 

\item \textbf{Zero-Day Attacks Handling}: Attacks evolve at a pace that datasets are not currently coping with. New dataset generation techniques are needed. If the process of generating datasets and making them publicly available is made more efficient, IDS models can be quickly updated and re-trained to cope with the changes.

\item \textbf{Special Purpose Datasets}: There are a limited number of available datasets serving special purpose IDSs. For example, publicly available datasets for IoT, SCADA and Tor networks are currently insufficient.

\item \textbf{Dataset Outlook}: Rapid advances in networking and associated technologies require a shift in dataset generation paradigm. Emerging technologies, such as Blockchain, Software Defined Network~(SDN), Network Function Virtualisation~(NFV), Big-Data, and their associated threats are currently not covered within available datasets. Yielding the dataset generation following trends in technologies~\cite{8274922, 10.1145/3309194.3309199, derhab2019blockchain}.

\end{itemize}

\subsection{Recommendations}
Guided by the critical impact of datasets on the evolvement of IDSs and the importance of robust and accurate IDS models, the following recommendations help build the next generation IDSs. The research direction should focus equally on building complex models for IDSs and gathering/generating data that represent benign and attack scenarios accurately. This will result in IDSs suitable for real-life deployments.  
\begin{itemize}
    \item \textbf{ML-First Vs Data-First}: As discussed in Section~\ref{sec:dataset}, obtaining valid, representative, and accurate data should be considered as the primary focus of research for the creation of IDSs. Building IDSs based on skewed and biased data only produces models unfit for exploitation, hence, Data-First models must be considered before ML-First.
     
    \item \textbf{Using precise evaluation metrics}: As discussed in Section~\ref{sec:IDS}, metrics -~other than accuracy~- should be considered to precisely reflect an IDS performance. For example, FP and Recall should be reported. Furthermore, the geometric mean should be used with imbalanced datasets, as well as, networking metrics such as throughput. 
    Conventional ML models report loss and accuracy by default unless other parameters are defined. Relying on the recorded loss and accuracy without measuring proposed evaluation metrics may result in misleading assessments of the overall IDS efficiency.
 
    \item \textbf{Introduce modular and extendable datasets}: As aforementioned, special purpose datasets are demanded, either to cover bespoke networks and architectures (e.g. IoT, SCADA, Tor, etc.) or to introduce new and zero-day attacks. 
    To increase the impact of datasets, they are required to be easily extendable and capable of integrating with other datasets. As a result, datasets would be adaptable to the continuous network changes. Also, dataset generation could be rendered in the IDS pipeline, therefore, not requiring the generation of a new dataset with every introduced change. To this end, anomaly based IDSs could be trained to use advanced ML techniques to identify new and zero-day attacks.
     
    \item \textbf{Standardize attack dataset generation/collection method}: One of the main challenges forcing researchers to work with outdated datasets is the lack of documentation associated with newly available datasets. Moreover, publishing raw packet data, not only the computed features, is needed to expand the use of datasets. One of the ways to generate datasets relies on $\alpha$ (using descriptive language to describe attacks) and $\beta$ (using behavior and statistical measures to describe attacks) profiles, as described by Shiravi~\textit{et al.}~\cite{RefWorks:doc:5cd437dce4b06db43e0599cc}.
    Ring~\textit{et al.}~\cite{RefWorks:doc:5cd94fbfe4b0de98174c4e1c} recommends being careful with anonymization and choosing which fields to be discarded.
    
    Privacy is depicted as one of the main obstacles against the collection of new attacks data. Furthermore, the lack of standard tools for data collection, anonymization, documentation and publication demand researchers to use their own tailored methods.  
    
    \item \textbf{Introduce models to inject realistic attacks}: Flow-Level Anomaly Modeling Engine~\textit{FLAME}~\cite{RefWorks:doc:5cd4264ce4b0ddd53ff48cfd} was one of the first tools to inject attacks that leave traces into network flows. \textit{ID2T}~\cite{RefWorks:doc:5cd42744e4b029a7217d0f41, RefWorks:doc:5cd4229ae4b004a2de355491, RefWorks:doc:5cd42318e4b073905aebe281} is a proposed flexible model to inject scenarios to existing datasets.
    
    The absence of thorough documentation of datasets makes it harder to map one dataset to another, rendering it impractical to add new attacks to existing datasets. Moreover, assuming that there exists a standard method to export realistic traffic, there isn't enough information about how to inject newly collected data to existing ones. Furthermore, the existing traffic injection proposals are limited to a single usage/prototype.
    
    \item \textbf{Generated dataset resilience}:
    To ensure dataset resilience, variations of the dataset should be generated. This could include variation of attack scenarios, different attacks, diverse benign traffic load (different time of day load). This, with other measures, would guarantee an extended dataset lifetime. Moreover, the complexity of  dataset variation generation should be kept minimal.
    It is key for a dataset to be provided in a raw state to allow researchers to make a choice between using stateless or stateful analysis.  If a dataset is provided only in a pre-processed form, researchers may lose the ability to use stateless or stateful detection methods and hinder the detection and accuracy of their algorithms.
    
    \item \textbf{Leveraging network monitoring to create real traffic}: Since most of the available benchmark datasets lack real-life properties, new datasets generation should benefit from network monitoring to generate realistic background traffic~\cite{bhuyan2015towards}. Furthermore, if real traffic could not be included in the dataset due to privacy concerns, real traffic should act as the ground truth for further traffic generation/simulation. 
    Moreover, for special purpose networks, relying on released  IoT/Critical infrastructure network architecture case studies should act as a guidance for network simulation. This should reduce the gap in terms of accurate realistic datasets.
    Different validation approaches should be used for IDS, however, with the advancement of IDS research, realistic experiments should be the main focus as demonstrated in~\cite{BEHAL20167}.
    
    \item \textbf{Dataset Validation:} Newly generated datasets should be validated based on network traffic validation techniques. Moln{\'a}r~\textit{et al.}~\cite{molnar2013validate} list the various metrics that could be used for this purpose. Furthermore, the similarities between real and synthetic traffic should also be evaluated as proposed in~\cite{hernandez2004real, cao2004stochastic}.
    
    \item \textbf{Updated Threats Taxonomy}: The networking threat taxonomy presented at this work aims at helping create datasets that cover a wider range of attacks.
    Maintaining the taxonomy in a timely manner will keep it an up-to-date reference for future IDSs research. Furthermore, the taxonomy is made available for public contribution through a GitHub repository to encourage contributions from other researchers to extend, revise and update it.

\end{itemize}

\noindent
While these recommendations might appear trivial at first, the majority of recent and old datasets proposed online do not conform to these guidelines as demonstrated within the previous sections. Hence, through this section, we provided recommendations for future datasets to follow ensuring the creation/generation of usable/accurate datasets.
\section{Conclusion}
\label{sec:Conclusion}
This research aims at tackling the problem of having a generic taxonomy for network threats. A proposed taxonomy is presented for categorizing network attacks based on the source, OSI model layer and whether the threat is active or passive. 
The prominent IDS research over the past decade (2008 - 2020) is analyzed. The analysis results in three main findings. First, benchmark datasets lack real-world properties and fail to cope with constant changes in attacks and network architectures, thus, limiting the performance of IDS.
Second, we present a taxonomy of tools and associated attacks, and demonstrate that current IDS research only covers around 33.3\% of threats presented in the taxonomy. Third, we highlight that - whilst ML is used by 97.25\% of the examined IDS - ANN, k-means and SVM represent the majority of the algorithms used. While these algorithms present outstanding results, we also highlight that these results are obtained on outdated datasets and, therefore, not representative of real-world architectures and attack scenarios. 

Finally, the network threat taxonomy and the attacks and associated tool taxonomy are open-sourced and available through GitHub\footnote{https://github.com/AbertayMachineLearningGroup/network-threats-taxonomy}, allowing both security and academic researchers to contribute to the taxonomy and ensure its relevance in the future.
\appendices
\section{}
\label{app:ids-comparison}
\setcounter{table}{0}
\renewcommand{\thetable}{A.\arabic{table}}
In this appendix, Table~\ref{tab:appendix} shows the prominent research over the past decade~(2008 - 2020) used in the analysis presented in Section~\ref{sec:recent-ids}. Each row represents one manuscript, highlighting the dataset and algorithms used within the research, alongside the attacks that the IDS is capable of detecting.
Table~\ref{tab:attacks-remarks} summarizes the publication year and attacks remarks for datasets discussed in Section~\ref{sec:dataset}.

\onecolumn
\tablefirsthead{%
     \hline 
     \rowcolor{gray!20}
      \textbf{Year} & \textbf{Dataset} & \centering\textbf{Used Algorithms} & \centering\textbf{Detected Attacks} & \textbf{Ref} \\ 
      }
\tablehead{%
     \hline
     \rowcolor{gray!20} 
     \textbf{Year} & \textbf{Dataset} & \centering\textbf{Used Algorithms} & \centering\textbf{Detected Attacks} & \textbf{Ref} \\}
\tabletail{%
\hline 
   \multicolumn{5}{r}{{Continued \ldots}} \\
}
\tablelasttail{
\hline
}
\tablecaption{Over A Decade of Intrusion Detection Systems~(2008 - 2020)}
\label{tab:appendix}
\begin{supertabular}{| >{
\centering\arraybackslash}m{0.04\linewidth} 
            | m{0.15\linewidth}
            | m{0.32\linewidth}
            | m{0.32\linewidth}
            | c|}
\hline 

2008 & KDD-99 & - Tree Classifiers \newline - Bayesian Clustering & Probing, DoS, R2L, U2R & \cite{RefWorks:doc:5adf2585e4b0c16216f3babf} 

\\ \hline  

2008 &  KDD-99 & - Parzen Classifier \newline - v-SVC \newline - k-means & Probing, DoS, R2L, U2R & \cite{RefWorks:doc:5adf22b7e4b066d2d902bc8d}

\\ \hline  

2008 & 1) PIERS \newline 2) Emergency Department Dataset \newline 3) KDD-99 & APD \newline - Bayesian Network Likelihood \newline - Conditional Anomaly Detection \newline - WSARE & 1) Illegal activity in imported containers \newline 2) Anthrax  \newline 3) DoS and R2L & \cite{RefWorks:doc:5ae30b22e4b0cc4b01647c4d}

\\ \hline  

2008 &  KDD-99 & - AdaBoost & Probing, DoS, R2L, U2R& \cite{RefWorks:doc:5adf24a5e4b09318b7179c0b}

\\ \hline  

2009 &  KDD-99 & - ABC \newline - Fuzzy Association Rules & Probing, DoS, R2L, U2R& \cite{RefWorks:doc:5adf20d4e4b066d2d902bc4e}

\\ \hline  

2009 & Collected transactions dataset  & - Fuzzy Association Rules & Credit Card Fraud & \cite{RefWorks:doc:5adf2e42e4b0e00594a4ca47}

\\ \hline  

2009 &  KDD-99 & - Genetic-based & Probing, DoS, R2L, U2R& \cite{RefWorks:doc:5adf5686e4b0c16216f3c546}

\\ \hline  

2009 &  KDD-99 & - C4.5 & Probing, DoS, R2L, U2R& \cite{RefWorks:doc:5adf2fa8e4b0a553e075025d}

\\ \hline  

2009 &  KDD-99 & BSPNN using: \newline - AdaBoost \newline - Semi-parametric NN & Probing, DoS, R2L, U2R& \cite{RefWorks:doc:5adf2996e4b09318b7179d65}

\\ \hline  

2009 & 1999 DARPA &
\begin{tabular}{@{}m{0.48\linewidth}m{0.48\linewidth}@{}}
- RBF & - Elman NN
\end{tabular} 
& Probing, DoS, R2L, U2R& \cite{RefWorks:doc:5adf2193e4b0cc4b01639d07}

\\ \hline  

2009 & 1999 DARPA & - SNORT \newline - Non-Parametric CUSUM \newline - EM based Clustering
 & 13 Attack Types & \cite{RefWorks:doc:5ae311d7e4b0a553e075c1d2}

\\ \hline  

2010 & KDD-99 & 
\begin{tabular}{@{}m{0.48\linewidth}m{0.48\linewidth}@{}}
FC-ANN based on: & \\
- Fuzzy Clustering & - ANN
\end{tabular}
  & Probing, DoS, R2L, U2R& \cite{RefWorks:doc:5ade07cce4b0f89694225437}

\\ \hline  

2010 &  KDD-99 & - Logistic Regression & Probing, DoS, R2L, U2R& \cite{RefWorks:doc:5adf5246e4b0cc4b0163a989}

\\ \hline  

2010 & KDD-99 &
\begin{tabular}{@{}m{0.48\linewidth}m{0.48\linewidth}@{}}
- FCM Clustering & - NN 
\end{tabular}
& Probing, DoS, R2L, U2R& \cite{RefWorks:doc:5ae07155e4b0f8969422cc73}

\\ \hline  

2011 &  Generated dataset & - OCSVM & - Nachi / Netbios scan \newline - DDoS UDP/ TCP flood \newline - Stealthy DDoS UDP flood \newline - DDoS UDP flood + traffic deletion Popup spam \newline - SSH scan + TCP flood & \cite{RefWorks:doc:5ae313b5e4b0a553e075c243}

\\ \hline  

2011  &  KDD-99 & 
\begin{tabular}{@{}m{0.48\linewidth}m{0.48\linewidth}@{}}
- AdaBoost & - NB \\
\end{tabular}
& Probing, DoS, R2L, U2R& \cite{RefWorks:doc:5adf2a4ce4b0c16216f3bb77}

\\ \hline  

2011 &  KDD-99 & - Genetic Algorithm \newline - Weighted k-NN & DoS / DDoS & \cite{RefWorks:doc:5ae075bfe4b029aa3ef8d3e7}

\\ \hline

2011  &  KDD-99 & 
\begin{tabular}{@{}m{0.48\linewidth}m{0.48\linewidth}@{}}
\multicolumn{2}{@{}l}{Genetic Fuzzy Systems based on:} \\
- Michigan & - Pittsburgh\\
- IRL & 
\end{tabular}
 & Probing, DoS, R2L, U2R& \cite{RefWorks:doc:5ae073a2e4b066d2d9030591}

\\ \hline  

2011 &   KDD-99 & \begin{tabular}{@{}m{0.48\linewidth}m{0.48\linewidth}@{}}
- DT & - Ripper Rule  \\
- BON & - RBF NN \\ 
\multicolumn{2}{@{}l}{- Bayesian Network}\\
- NB & 
\end{tabular}
 & - Probing \newline - DoS & \cite{RefWorks:doc:5ade0794e4b029aa3ef82ecf}

\\ \hline  

2011 &  KDD-99 & - K-means clustering \newline - SOM & Probing, DoS, R2L, U2R& \cite{RefWorks:doc:5ae0779ae4b066d2d9030662}

\\ \hline  

2011 &  KDD-99 & 
\begin{tabular}{@{}m{0.48\linewidth}m{0.48\linewidth}@{}}
- Rule-Based & - BON \\
- ART Network
\end{tabular}
  & Probing, DoS, R2L, U2R& \cite{RefWorks:doc:5ae075a9e4b0e00594a51133}

\\ \hline  

2011 &  KDD-99 &  - SVM  & Probing, DoS, R2L, U2R & \cite{RefWorks:doc:5adf2b3be4b0cc4b01639f93}

\\ \hline  

2011 &  KDD-99 &
\begin{tabular}{@{}m{0.48\linewidth}m{0.48\linewidth}@{}}
- K-Means & - NB
\end{tabular}
  & Probing, DoS, R2L, U2R& \cite{RefWorks:doc:5ae3179be4b0fe8e18a65c9a}

\\ \hline  

2012  & KDD-99 & 
\begin{tabular}{@{}m{0.48\linewidth}m{0.48\linewidth}@{}}
- Modified SOM & - k-means
\end{tabular}
 & Probing, DoS, R2L, U2R & \cite{RefWorks:doc:5ae07b66e4b066d2d903075b}

\\ \hline  

2012 & 1998 DARPA &  - SVM  & Attack and Non-Attack & \cite{RefWorks:doc:5adf547be4b066d2d902c888}

\\ \hline  

2012 &  1998 DARPA & 
\begin{tabular}{@{}m{0.48\linewidth}m{0.48\linewidth}@{}}
\multicolumn{2}{@{}l}{ELMs:} \\
- Basic & - Kernel-Based
\end{tabular}
& Probing, DoS, R2L, U2R& \cite{RefWorks:doc:5ae31eebe4b066d2d903986d}

\\ \hline  

2012 & 1998 DARPA &  - SVDD  & U2R & \cite{RefWorks:doc:5ae31b0de4b07da0d123f927}

\\ \hline  

2012 &  KDD-99 &  - Hidden NB  & Probing, DoS, R2L, U2R& \cite{RefWorks:doc:5ae1b485e4b029aa3ef91f7a}

\\ \hline  

2012 &  KDD-99 & 
\begin{tabular}{@{}m{0.48\linewidth}m{0.48\linewidth}@{}}
- SVM &  - DT \\
- SA & 
\end{tabular}
 & Probing, DoS, R2L, U2R& \cite{RefWorks:doc:5ade0607e4b0cc4b016355c6}

\\ \hline  

2012 & KDD-99 &  
\begin{tabular}{@{}m{0.48\linewidth}m{0.48\linewidth}@{}}
\multicolumn{2}{@{}l}{Ensemble DTs: }\\
- Decision Stump  & - C4.5 \\
- NB Tree &  - RF \\
- Random Tree & \\
\multicolumn{2}{@{}l}{-  Representative Tree model}
\end{tabular}
 & Probing, DoS, R2L, U2R& \cite{RefWorks:doc:5adf2bfce4b0e00594a4c9c2}

\\ \hline  

2012  & KDD-99 &  
\begin{tabular}{@{}m{0.48\linewidth}m{0.48\linewidth}@{}}
- K-means & - SVM\\
- Ant Colony
\end{tabular}
& Probing, DoS, R2L, U2R& \cite{RefWorks:doc:5ae07a97e4b029aa3ef8d5ae}

\\ \hline  

2013 & KDD-99  & - Fuzzy C means \newline - Fuzzy NN / Neurofuzzy \newline - RBF SVM & Probing, DoS, R2L, U2R& \cite{RefWorks:doc:5ae07bf2e4b0e00594a51263}

\\ \hline  

2013 & NSL-KDD & - Fuzzy Clustering NN & Probing, DoS, R2L, U2R& \cite{RefWorks:doc:5ae07cafe4b0cc4b0163e91c}

\\ \hline  

2013 & KDD-99 & 
\begin{tabular}{@{}m{0.48\linewidth}m{0.48\linewidth}@{}}
- K-means & - NN MLP
\end{tabular}
 & Probing, DoS, R2L, U2R & \cite{RefWorks:doc:5ae07d8ee4b0fe8e18a5d76f}

\\ \hline  

2013  & KDD-99 & 
\begin{tabular}{@{}m{0.48\linewidth}m{0.48\linewidth}@{}}
- FFNN & - ENN \\
- GRNN & - PNN \\
- RBNN &
\end{tabular}
 & Probing, DoS, R2L, U2R & \cite{RefWorks:doc:5adf55fbe4b0a553e0750b6a}

\\ \hline  

2013 & DARPA 2000 & APAN using: \newline - Markov Chain \newline - Kmeans Clustering & DDoS & \cite{RefWorks:doc:5addfea8e4b0c16216f361c6}

\\ \hline  

2013 & ISCX 2012 & KMC+NBC \newline - K-Means Clustering \newline - NB Classifier & Normal and Attack & \cite{RefWorks:doc:5ae3354de4b0fe8e18a660ea}

\\ \hline  

2013 & Bank's Credit Card Data  &  - DT & Fraud & \cite{RefWorks:doc:5adf54f7e4b066d2d902c898}

\\ \hline  

2013 & KDD-99 & Two variants of GMDH: \newline - Monolithic \newline - Ensemble-based & Probing, DoS, R2L, U2R& \cite{RefWorks:doc:5addfe77e4b0fe8e18a54265}

\\ \hline  

2013  & Simulated dataset & - Non-Parametric CUSUM & Jamming  & \cite{RefWorks:doc:5ae1f5f1e4b066d2d9035b91}

\\ \hline  

2014 & - KDD-99 & - ELM  & Probing, DoS, R2L, U2R & \cite{RefWorks:doc:5bbb4cd3e4b0dfeb9535642b}

\\ \hline  

2014  & - KDD-99 \newline - NSL-KDD & ANN-Bayesian Net-GR ensemble: \newline - ANN \newline - Bayesian Net with GR feature selection & Probing, DoS, R2L, U2R& \cite{RefWorks:doc:5adf2cade4b0c16216f3bbf8}

\\ \hline  

2014 & NSL-KDD & 
\begin{tabular}{@{}m{0.48\linewidth}m{0.48\linewidth}@{}}
-  One-class SVM & - C4.5 DT
\end{tabular}
 & - & \cite{RefWorks:doc:5addf22ae4b0cc4b01634b92}

\\ \hline  

2014  & KDD-99 & - K-medoids & Probing, DoS, R2L, U2R& \cite{RefWorks:doc:5ae1d13de4b029aa3ef92a7a}

\\ \hline  

2014  & KDD-99 & 
\begin{tabular}{@{}m{0.48\linewidth}m{0.48\linewidth}@{}}
- SVM & - CSOACN
\end{tabular}
& Probing, DoS, R2L, U2R& \cite{RefWorks:doc:5addef77e4b0c16216f35baf}

\\ \hline

2014 & NSL-KDD & - AIS~(NSA, CSA, INA) & Normal and abnormal & \cite{AFZALISERESHT2014286}

\\ \hline  

2015 & KDD-99 & - DT \newline - CFA (Feature Selection) & Probing, DoS, R2L, U2R& \cite{RefWorks:doc:5addee2be4b029aa3ef82288}

\\ \hline  

2015 & gureKddcup6percent & - SVM & R2L & \cite{RefWorks:doc:5ae1f407e4b0e00594a58d29}

\\ \hline  

2015 & KDD-99 & 
\begin{tabular}{@{}m{0.48\linewidth}m{0.48\linewidth}@{}}
- K-means & - k-NN 
\end{tabular}
& Probing, DoS, R2L, U2R& \cite{RefWorks:doc:5adde6f7e4b029aa3ef8207d}

\\ \hline  

2015 & KDD-99 & - Weighted ELM & Probing, DoS, R2L, U2R& \cite{RefWorks:doc:5ae32072e4b0a553e075c3fc}

\\ \hline

2015 & GureKddcup & - AIS (R-chunk) & Normal and abnormal & \cite{7397274}

\\ \hline  

2016 & KDD-99 & 
\begin{tabular}{@{}m{0.48\linewidth}m{0.48\linewidth}@{}}
- PCA & - Fuzzy PCA  \\
- k-NN &
\end{tabular}
 & Probing, DoS, R2L, U2R& \cite{RefWorks:doc:5ae20368e4b0a553e0758c04}

\\ \hline  

2016  & NSL-KDD & 
\begin{tabular}{@{}m{0.48\linewidth}m{0.48\linewidth}@{}}
- PCA & - SVM \\
- MLP &  - C4.5 \\
- NB 
\end{tabular}
& Probing, DoS, R2L, U2R& \cite{RefWorks:doc:5ae1ff0fe4b09318b7184781}

\\ \hline  

2016 &  Simulated dataset & - ANN  & DoS/DDoS  & \cite{RefWorks:doc:5afd8f7ae4b013f23759cb5e}

\\ \hline  

2016 &  Generated dataset using httperf & - Mapping & - SQL Injection \newline - XSS & \cite{RefWorks:doc:5ae205bae4b066d2d903612a}

\\ \hline  

2016 & KDD-99 & 
\begin{tabular}{@{}m{0.48\linewidth}m{0.48\linewidth}@{}}
- SVM &  - PCA
\end{tabular}
& - Normal and Attack & \cite{RefWorks:doc:5ae20350e4b0cc4b01644d87}

\\ \hline 
2016 & NSL-KDD & 
\begin{tabular}{@{}m{0.48\linewidth}m{0.48\linewidth}@{}}
- AIS (NSA-GA) & - SVM \\
- NB & - DT (J48)
\end{tabular}
& Normal and abnormal & \cite{7545821}
\\ \hline  

2017 & - Kyoto2006+ \newline - NSL-KDD & - Forked VAE \newline - Unsupervised deep NN & Probing, DoS, R2L, U2R & \cite{RefWorks:doc:5bbb5223e4b05d07c954cb9a} 

\\ \hline  

2017 & KDD-99 & 
\begin{tabular}{@{}m{0.48\linewidth}m{0.48\linewidth}@{}}
- Binary PSO & - k-NN
\end{tabular}
 & Probing, DoS, R2L, U2R& \cite{RefWorks:doc:5ae1f962e4b0f896942310c5}

\\ \hline  

2017 & KDD-99 &
\begin{tabular}{@{}m{0.48\linewidth}m{0.48\linewidth}@{}}
- R-tree & - k-NN \\
- K-means & - SVM
\end{tabular}
 & Probing, DoS, R2L, U2R & \cite{RefWorks:doc:5af0545ce4b0163ac8ff49f7} 

\\ \hline  

2017  & Generated dataset & 
\begin{tabular}{@{}m{0.48\linewidth}m{0.48\linewidth}@{}}
- GPU-based ANN & - BON
\end{tabular}
 & Normal and Attack & \cite{RefWorks:doc:5ae1ef4be4b0cc4b01644975}

\\ \hline  

2017 & NSL-KDD & - DL RNN & Probing, DoS, R2L, U2R & \cite{RefWorks:doc:5ae33844e4b029aa3ef96c92}

\\ \hline  

2017  & NSL-KDD &
\begin{tabular}{@{}m{0.48\linewidth}m{0.48\linewidth}@{}}
- K-means & - NB \\
- Information Gain &
\end{tabular}
 & Probing, DoS, R2L, U2R& \cite{RefWorks:doc:5ae1dc89e4b066d2d9035498}

\\ \hline  

2017 & UNB-CIC & 
\begin{tabular}{@{}m{0.48\linewidth}m{0.48\linewidth}@{}}
- ANN & - SVM 
\end{tabular}
 & nonTor Traffic & \cite{RefWorks:doc:5cdeb6e3e4b09ae3c887080c}

\\ \hline  

2017 & KDD-99 & - Polynomial Feature Correlation & - DoS & \cite{RefWorks:doc:5ae1fd8fe4b07da0d123d8cd}

\\ \hline  

2017  & KDD-99  & 
\begin{tabular}{@{}m{0.48\linewidth}m{0.48\linewidth}@{}}
- PCA & - k-NN \\
\multicolumn{2}{@{}l}{- Softmax Regression}
\end{tabular}
  & Probing, DoS, R2L, U2R & \cite{RefWorks:doc:5af05725e4b0b5a912ed9311}

\\ \hline  

2017 & KDD-99 & Optimized Backpropagation by Conjugate Gradient algorithm \newline - Fletcher Reeves \newline - Polak Ribiere \newline - Powell Beale & Probing, DoS, R2L, U2R & \cite{RefWorks:doc:5ae1ed1ce4b0a553e07587c4}

\\ \hline  

2018 & KDD-99 & - Kernel Clustering & Probing, DoS, R2L, U2R & \cite{RefWorks:doc:5af0521fe4b0c3f002684cfc}

\\ \hline  

2018 & Simulated Dataset & 
\begin{tabular}{@{}m{0.48\linewidth}m{0.48\linewidth}@{}}
- MLP & - SVM \\
- J48 & - NB \\
- Logistic & - RF \\
\multicolumn{2}{@{}l}{Features Selection: } \\
- BFS-CFS & - GS-CFS 
\end{tabular}
& Individual and Combination Routing Attacks: \newline - Hello Flood \newline - Sinkhole \newline - Wormhole & \cite{RefWorks:doc:5ae1e080e4b0a553e07585e9}

\\ \hline  

2018 & KDD-99 & 
\begin{tabular}{@{}m{0.48\linewidth}m{0.48\linewidth}@{}}
- FLN & - PSO
\end{tabular}
 & Probing, DoS, R2L, U2R & \cite{RefWorks:doc:5af059cfe4b02abf496dd40a}

\\ \hline  

2018 &  NSL-KDD \newline - UNSW-NB15 & - Deep Auto-Encoder \newline - ANN & Probing, DoS, R2L, U2R & \cite{RefWorks:doc:5b0e81c9e4b01f2c3e37bf75}

\\ \hline  

2018 & -KDD-99 \newline -NSL-KDD & 
\begin{tabular}{@{}m{0.48\linewidth}m{0.48\linewidth}@{}}
- DL & - NDAE \\
- Stacked NDAEs &
\end{tabular}
 & Probing, DoS, R2L, U2R& \cite{RefWorks:doc:5ae339c4e4b0e00594a5c5cc}

\\ \hline  

2018 &  KDD-99 & 
\begin{tabular}{@{}m{0.48\linewidth}m{0.48\linewidth}@{}}
- KFRFS & - NB \\
- IBK & - AdaBoost \\ 
- MFNN & - SMO \\
- RF &
\end{tabular}
  &  Probing, DoS, R2L, U2R & \cite{8491578} 
\\ \hline

2018 & NSL-KDD & - AIS~(NSA, CSA) & Normal and abnormal & \cite{8405726} 

\\ \hline

2018 & - KDD-99 \newline - CAIDA’07/08 \newline - Generated traffic & - AIS & DoS & \cite{VIDAL201894} 

\\ \hline

2019 & NSL-KDD & 
\begin{tabular}{@{}m{0.48\linewidth}m{0.48\linewidth}@{}}
- NB  &  - RF  \\
- ANN & - SVM \\
- BayesNet & \\
\multicolumn{2}{@{}l}{- DT (Enhanced J48, J48, ADTree, } \\ \multicolumn{2}{@{}l}{DecisionStump, RandomTree,} \\
\multicolumn{2}{@{}l}{SimpleCart)}
\end{tabular}
 & Probing, DoS, R2l, U2R & \cite{Aljawarneh2019} 
\\ \hline

2019 & KDD-99 & 
\begin{tabular}{@{}m{0.48\linewidth}m{0.48\linewidth}@{}}
- DT & \\
\multicolumn{2}{@{}l}{- SVM~(least square)} \\
\multicolumn{2}{@{}l}{Feature Selection:}\\ 
- FGLCC &  - CFA 
\end{tabular}
  & Probing, DoS, R2l, U2R & \cite{MOHAMMADI201980} 
\\ \hline 
 
2019 & - UNSW-NB15 \newline - CICIDS2017 & 
\begin{tabular}{@{}m{0.48\linewidth}m{0.48\linewidth}@{}}
- Deep FFNN & - RF \\
\multicolumn{2}{@{}l}{- Gradient Boosting Tree}
\end{tabular}
 & - Fuzzers, Analysis, Backdoors, DoS, Exploits, Generic, Reconnaissance, Shell-Code \& Worms \newline - DoS, DDoS, Web-based, Brute force, Infiltration, Heartbleed, Bot \& Scan & \cite{10.1145/3299815.3314439} 

\\ \hline
2019 & - ISCX 2012 \newline - NSL-KDD \newline - Kyoto2006+ &
\begin{tabular}{@{}m{0.48\linewidth}m{0.48\linewidth}@{}}
- IG & - PCA \\
- SVM & - IBK \\
- MLP
\end{tabular}
 & - Normal and Attack \newline -  Probing, DoS, R2l, U2R & \cite{SALO2019164} 
\\ \hline

2019 & - KDD-99 \newline - NSL-KDD \newline - UNSW-NB15 \newline - Kyoto2006+ \newline - WSN-DS \newline - CICIDS2017 & 
\begin{tabular}{@{}m{0.48\linewidth}m{0.48\linewidth}@{}}
\multicolumn{2}{@{}l}{- Deep NN} \\ 
\multicolumn{2}{@{}l}{- Logistic Regression} \\
- NB & - k-NN \\ 
- SVM & - DT \\
- AB & - RF 
\end{tabular}
 & - Probing, DoS, R2l, U2R  \newline - 4 DoS attacks (Blackhole, Grayhole, Flooding \& Scheduling) \newline - Fuzzers, Analysis, Backdoors, DoS, Exploits, Generic, Reconnaissance, Shell-Code \& Worms \newline - DoS, DDoS, Web-based, Brute force, Infiltration, Heartbleed, Bot \& Scan & \cite{8681044} 
\\ \hline

2019 & - NSL-KDD & 
\begin{tabular}{@{}m{0.48\linewidth}m{0.48\linewidth}@{}}
\multicolumn{2}{@{}l}{- Genetic Algorithms}\\
\multicolumn{2}{@{}l}{- Kernel ELM} \\
- DT & - k-NN \\
- MLP & - SVM
\end{tabular}
 & Probing, Dos, R2L, U2R & \cite{Ghasemi2019}
\\ \hline

2020 & Generated dataset & 
\begin{tabular}{@{}m{0.48\linewidth}m{0.48\linewidth}@{}}
- AdaBoost & - Decision Stump \\
- J48 & - RF \\
- SVM & - MLP \\
- NB & - BayesNet
\end{tabular}
 & - DDoS & \cite{10.1007/978-981-13-7564-4_5}

\\ \hline
2020 & NSL-KDD  & - Deep NN & Probing, DoS, R2l, U2R  & \cite{10.1007/978-981-15-0637-6_40}
\\ \hline 

2020 &  - KDD-99 &
\begin{tabular}{@{}m{0.48\linewidth}m{0.48\linewidth}@{}}
- Ontology & - NB \\
- DT & - RF
\end{tabular}
& Probing, Dos, R2L, U2R & \cite{sarnovsky2020hierarchical}
\\ \hline 

2020 & Generated dataset & - Local Outlier Factor \newline - Isolation Forest &  Port Scanning, HTTP \& SSH Brute
Force \&  SYN Flood & \cite{8976157}

\\ \hline
\end{supertabular}
\vspace{3mm}
Where: \\
\begin{tabular}{l l}
* 	ABC: Association Based Classification	& * 	IG: Information Gain	\\
* 	AdaBoost : Adaptive Boosting	& * 	INA: Immune Network Algorithms	\\
* 	AIS: Artificial Immune System	& * 	IRL: Iterative Rule Learning	\\
* 	ANN: Artificial Neural Network	& * 	KFRFS: Kernel-based Fuzzy-Rough Feature Selection	\\
* 	APAN: Advanced Probabilistic Approach for Network-based IDS	& * 	k-NN: k-Nearest Neighbors	\\
* 	APD: Anomaly Pattern Detection	& * 	MFNN: Multi-Functional Nearest-Neighbour	\\
* 	ART: Adaptive Resonance Theory	& * 	MLP: Multi-Layer Perceptron	\\
* 	BFS-CFS: Best First Search with Correlation Features Selection	& * 	NB: Na{\"i}ve Bayes	\\
* 	BON: Back-Propagation Network	& * 	NDAE: Non-Symmetric Deep Auto-Encoder	\\
* 	BSPNN: Boosted Subspace Probabilistic Neural Network	& * 	NN: Neural Network	\\
\end{tabular}

\begin{tabular}{l l}
* 	CFA: CuttleFish Algorithm	& * 	NSA: Negative Selection Algorithm	\\
* 	CSA: Clonal Selection Algorithm	& * 	OCSVM: One Class Support Vector Machine	\\
* 	CSOACN: Clustering based on Self-Organized Ant Colony Network 	& * 	PCA: Principal Component Analysis	\\
* 	CUSUM: CUmulative SUM	& * 	PNN: Probabilistic Neural Network	\\
* 	DL: Deep Learning	& * 	PSO: Particle Swarm Optimization	\\
* 	DoS: Denial of Service	& * 	R2L: Remote to Local	\\
* 	DT: Decision Tree 	& * 	RBF: Radial Basis Function	\\
* 	ELM: Extreme Learning Machine	& * 	RBNN: Radial Basis Neural Network	\\
* 	ENN: Elman Neural Network	& * 	RF: Random Forest	\\
* 	FCM: Fuzzy C-Mean	& * 	RNN: Recurrent Neural Networks	\\
* 	FFNN: Feed Forward Neural Network 	& * 	SA: Simulated Annealing 	\\
* 	FGLCC: Feature Grouping based on Linear Correlation Coefficient	& * 	SOM: Self-Organizing Map	\\
* 	FLN: Fast Learning Network	& * 	SVDD: Support Vector Data Description	\\
* 	GMDH: Group Method for Data Handling	& * 	SVM: Support Vector Machine	\\
* 	GR: Gain Ratio	& * 	U2R: User to Root	\\
* 	GRNN: Generalized Regression Neural Network	& * 	VAE: Variational Auto-Encoder	\\
* 	GS-CFS: Greedy Stepwise with Correlation Features Selection	& * 	WSARE: What’s Strange About Recent Events	\\
* 	IBK: Instance Based Learning	& * 	XSS: Cross Site Scripting	\\

\end{tabular}

\begin{table}[!th]
    \caption{Datasets Attacks Remarks}
    \begin{center}
    \begin{tabular}{
    |P{0.20\linewidth}  
    |P{0.30\linewidth}
    |P{0.50\linewidth}| }
    \hline
    \rowcolor{gray!20}
     \textbf{Dataset Name}  & \textbf{Institute} & \textbf{Attacks Remarks}\\
     \hline 
     \rowcolor{gray!20}
     \multicolumn{3}{|c|}{General Purpose Networks} \\
     \hline
     ADFA-IDS 2017 \cite{RefWorks:doc:5cd94680e4b0661506bd7d82, RefWorks:doc:5b06d3c3e4b00f93d242857a}  & Australian Defense Force Academy & - \\ \hline
     
     Booters \cite{RefWorks:doc:5cd970fde4b048754198ca22}  & University of Twente, SURFnet, Federal University of Rio Grande do Sul & 9 DDoS attacks \\ \hline
     
     Botnet dataset \cite{RefWorks:doc:5b227bf6e4b0d1cffc0657b8} &   Canadian Institute for Cybersecurity~(CIC) & 7 Botnet types in training set and 16 in test set  \\ \hline
     
     CAIDA 2007 \cite{caida2007} & Center for Applied Internet Data Analysis & 1 hour of DDoS attack divided into 5-minute pcap files \\ \hline
     
     CIC DoS dataset \cite{RefWorks:doc:5b227c16e4b0dcaa93352175} & \multirow{3}{\linewidth}{\centering CIC} &  8 DoS attack traces \\ \cline{1-1} \cline{3-3}
     
     CICIDS2017 \cite{RefWorks:doc:5b227bb8e4b07f83f15ddb45} &  &  \multirow{2}{\linewidth}{\centering 4 DoS types, Infiltration Dropbox Download and Cool disk, 14 Port Scan types}\\ \cline{1-1}
     CICIDS2018 \cite{sharafaldin2018toward} &   & \\ \hline
     
      CTU-13 \cite{garcia2014empirical} & CTU University & 13 captures of botnet samples\\ \hline
      
      DARPA \cite{RefWorks:doc:5b0693fce4b0faf8604e2beb} & MIT Lincoln Laboratory & 17 DoS, 12 U2R, 15 R2L, 10 Probing and 1 Data \\  \hline
      DDoSTB~\cite{behal2016measuring} & Punjab Technical University \& SBS State Technical Campus & DDoS Testbed using emulated and real nodes \\ \hline
      ISCXIDS2012 \cite{RefWorks:doc:5b06d318e4b0faf8604e40ce} &  CIC & HTTP, SMTP, SSH, IMAP, POP3, and FTP Traffic \\ \hline
      
      KDD-99 \cite{RefWorks:doc:5b06959ce4b078503013ca8c}  & University of California & Covers 24 training attack types and 14 additional types in the test data \\ \hline
      TUIDS~\cite{bhuyan2015towards} & Tezpur University & (1)~TUIDS IDS dataset. (2)~TUIDS Scan dataset. (3)~TUIDS DDoS dataset (22 DDoS attack types)  \\ \hline
      NSL-KDD \cite{RefWorks:doc:5b227bd4e4b0d1cffc0657b0}  & CIC & Improvement of KDD'99 dataset  \\ \hline
     
      STA2018 \cite{amjad_al_tobi_2018}& University of St Andrews & Transformation of UNB ISCX (contains 550 features) \\ \hline 
    
     Unified Network Dataset \cite{turcotte17} & Los Alamos National Laboratory & ~90 days of Network and Host logs \\ \hline
    
    Waikato~\cite{waikato} & RIPE Network Coordination Center & - \\ \hline
     \rowcolor{gray!20}
     \multicolumn{3}{|c|}{Special Purpose Networks} \\
     \hline
     4SICS ICS \cite{SCADAICS9:online} & Netresec & - \\ \hline
     Anomalies Water System \cite{RefWorks:doc:5ad0a812e4b0e18303dc2c3e, 10.1007/978-3-030-12786-2_1} & French Naval Academy & 15 different real situations covering cyber-attacks~(DoS \& Spoofing), breakdown (Sensor Failure \& Wrong connection),  sabotage (Blocked Measures \& Floating Objects)  \\ \hline
      
     Bot-IoT \cite{koroniotis2018towards} & The center of UNSW Canberra Cyber & Attacks include DoS/DDoS, OS and Service Scan, Keylogging and Data Exfiltration \\ \hline

     IoT devices captures \cite{miettinen2017iot}   & Aalto University &  represents that data of 31 smart home IoT devices of 27 different types \\ \hline
      
      Tor-nonTor dataset \cite{RefWorks:doc:5b22790ee4b032ce1ef1ee04} & \multirow{2}{\linewidth}{\centering CIC} & 7 traffic categories (Browsing, Email, Chat, Audio/Video-Streaming,  FTP, VoIP, P2P) \\ \cline{1-1} \cline{3-3}
     
      VPN-nonVPN dataset \cite{RefWorks:doc:5b22787ce4b034677a1ee449} &  & 14 traffic categories (VPN-VOIP, VPN-P2P, etc.) covering Browsing, Email, Chat, Streaming, File Transfer, VoIP, TraP2P \\ \hline 
     \rowcolor{gray!20}
     \multicolumn{3}{|c|}{Mobile Applications} \\
     \hline
     Android Adware and General Malware Dataset \cite{RefWorks:doc:5b227673e4b03aeb784ad16a}  & \multirow{2}{\linewidth}{\centering CIC} & 1900 application (Adware, General Malware and Benign) \\ \cline{1-1} \cline{3-3} 
     
     Android Botnet dataset \cite{RefWorks:doc:5b227757e4b0dcaa93351fbf}  &  & 1929 Botnet samples covering 14 families (AnserverBot, Bmaster, DroidDream, Geinimi, MisoSMS, NickySpy, Not Compatible, PJapps, Pletor, RootSmart, Sandroid, TigerBot, Wroba, Zitmo \\ \hline
     
     Android Malware Genome \cite{zhou2012dissecting} & North Carolina State University & More than 1,200 malware samples \\ \hline
     
     AndroMalShare \cite{AndroMal22:online} & Botnet Research Team and Xi'an Jiaotong University & More than 85,000 Android malware samples \\ \hline
     
     Kharon Malware Dataset \cite{kiss:hal-01300752}  & Kharon project & 7 malware deeply examined, 10 malware (one sample each) and 2 partially examined\\ \hline
    
    \end{tabular}
    \end{center}
    \label{tab:attacks-remarks}
\end{table}
\twocolumn

\bibliographystyle{IEEEtran}
\bibliography{8-bibliography}

\begin{IEEEbiography}[{\includegraphics[width=1in,height=1.25in,clip,keepaspectratio]{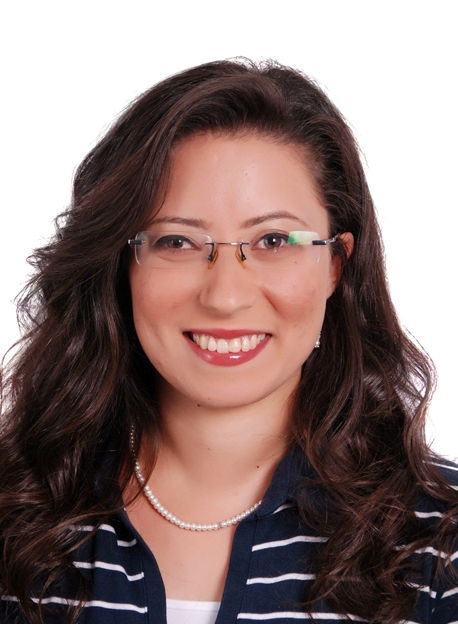}}]{Hanan Hindy} 
is a second year Ph.D. student at the Division of Cyber-Security at Abertay University, Dundee, Scotland. Hanan received her bachelor degree with honours (2012) and a masters (2016) degrees in Computer Science from the Faculty of Computer and Information Sciences at Ain Shams University, Cairo, Egypt. Her research interests include Machine Learning and Cyber Security. She is currently working on utilising deep learning for IDS.
\end{IEEEbiography}

\begin{IEEEbiographynophotoAlt}{David Brosset} received his master's degree in computer science from the university of south Brittany in 2003 and his Ph.D in computer science in 2008 from the Arts et Metiers, Paris. Since 2011, he is an associate professor of Computer Science at the French Naval Academy and he is involved in the chair of cyber defence of naval systems. His research concerns the domain of cyber security and in particular the cyber defence of critical systems on board.
\end{IEEEbiographynophotoAlt}

\begin{IEEEbiography}[{\includegraphics[width=1in,height=1.25in,clip,keepaspectratio]{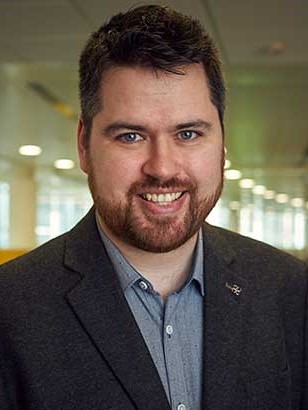}}]{Ethan Bayne} 
received a BSc degree in Computing and Networks; a MSc degree in Ethical Hacking and Computer Security; and a PhD degree in Digital Forensics from  Abertay University (Dundee, Scotland) in  2008,  2013,  and  2016,  respectively. He is currently a Lecturer in Cyber Security and Computer Science within the Department of Cyber Security at Abertay University. His current research interests include digital forensics, massively-parallel computation, pattern matching, machine learning and internet of things (IoT).
\end{IEEEbiography}

\begin{IEEEbiography}[{\includegraphics[width=1in,height=1.25in,clip,keepaspectratio]{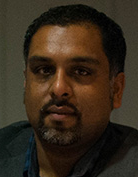}}]{Amar Seeam}
is a Senior Lecturer in Computer Science at Middlesex University Mauritius. He obtained his PhD from the University of Edinburgh in 2015. Other qualifications obtained by Amar include a BEng (Hons) in Mechanical Engineering (2003) and MSc in Information Technology (2004) conferred by the University of Glasgow as well as an MSc in System Level Integration from the University of Edinburgh (2005). His research interests include Simulation Assisted Control, Cybersecurity, Internet of Things and Building Information Modeling.
\end{IEEEbiography}

\begin{IEEEbiography}[{\includegraphics[width=1in,height=1.25in,clip,keepaspectratio]{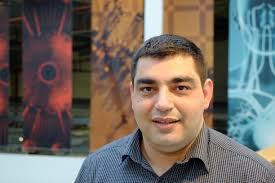}}]{Christos Tachtatzis} is a Senior Lecturer Chancellors Fellow in Sensor Systems and Asset Management, at the University of Strathclyde. He holds a BEng (Hons) in Communication Systems Engineering from University of Portsmouth in 2001, an MSc in Communications, Control and Digital Signal Processing (2002) and a PhD in Electronic and Electrical Engineering (2008), both from Strathclyde University. Christos has 12 years experience, in Sensor Systems ranging from electronic devices, networking, communications and signal processing. His current research interests lie in extracting actionable information from data using machine learning and artificial intelligence.
\end{IEEEbiography}

\begin{IEEEbiography}[{\includegraphics[width=1in,height=1.25in,clip,keepaspectratio]{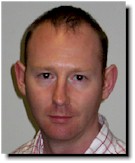}}]{Robert Atkinson}
received the B.Eng. (Hons.) degree in electronic and electrical engineering; the M.Sc. degree in communications, control, and digital signal processing; and the Ph.D. degree in mobile communications systems from the University of Strathclyde, Glasgow, U.K., in 1993, 1995, and 2003, respectively. He is currently a Senior Lecturer at the institution. His research interests include data engineering and the application of machine learning algorithms to industrial problems including cyber-security.
\end{IEEEbiography}

\begin{IEEEbiography}[{\includegraphics[width=1in,height=1.25in,clip,keepaspectratio]{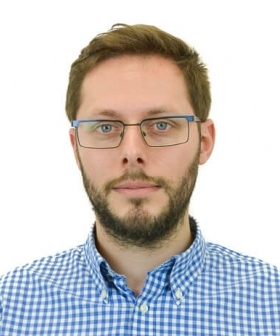}}]{Xavier Bellekens} received the bachelor’s degree from HeNam, Belgium, in 2010, the master’s degree in ethical hacking and computer security from the University of Abertay, Dundee, in 2012, and the Ph.D. degree in electronic and electrical engineering from the University of Strathclyde, Glasgow, in 2016. He is currently a Chancellor’s Fellow Lecturer with the Department of Electronic and Electrical Engineering, University of Strathclyde, where he has been working on cyber-security for critical infrastructures. Previously, he was a Lecturer in security and privacy with the Department of Cyber-Security, University of Abertay, where he led the Machine Learning for Cyber-Security Research Group. His current research interests include machine learning for cyber-security, autonomous distributed networks, the Internet of Things, and critical infrastructure protection.
\end{IEEEbiography}
\EOD

\end{document}